\documentclass[prd,aps,preprintnumbers, showpacs, nofootinbib,superscriptaddress,notitlepage, 10pt]{revtex4-1}
\usepackage{epsfig}
\usepackage{bm}
\usepackage{amssymb}
\usepackage{amsmath}
\usepackage{color}
\usepackage{subfigure}
\usepackage[colorlinks,
            linkcolor=blue,
            anchorcolor=black,
            citecolor=blue
            ]{hyperref}

\begin{document}

\title{Study of Isolated-photon and Jet Momentum Imbalance in $pp$ and $PbPb$ collisions}

\author{Lin Chen}

\affiliation{Key Laboratory of Quark and Lepton Physics (MOE) and Institute
of Particle Physics, Central China Normal University, Wuhan 430079, China}

\author{Guang-You Qin}

\affiliation{Key Laboratory of Quark and Lepton Physics (MOE) and Institute
of Particle Physics, Central China Normal University, Wuhan 430079, China}

\author{Lei Wang}
\affiliation{Key Laboratory of Quark and Lepton Physics (MOE) and Institute
of Particle Physics, Central China Normal University, Wuhan 430079, China}

\author{Shu-Yi Wei}
\affiliation{Centre de Physique Th\'eorique, \'Ecole Polytechnique, 
CNRS, Universit\'e Paris-Saclay, Route de Saclay, 91128 Palaiseau, France.}

\affiliation{Key Laboratory of Quark and Lepton Physics (MOE) and Institute
of Particle Physics, Central China Normal University, Wuhan 430079, China}

\author{Bo-Wen Xiao}
\affiliation{Key Laboratory of Quark and Lepton Physics (MOE) and Institute
of Particle Physics, Central China Normal University, Wuhan 430079, China}
\affiliation{Centre de Physique Th\'eorique, \'Ecole Polytechnique, 
CNRS, Universit\'e Paris-Saclay, Route de Saclay, 91128 Palaiseau, France.}

\author{Han-Zhong Zhang}
\affiliation{Key Laboratory of Quark and Lepton Physics (MOE) and Institute
of Particle Physics, Central China Normal University, Wuhan 430079, China}

\author{Ya-Qi Zhang}
\affiliation{Key Laboratory of Quark and Lepton Physics (MOE) and Institute
of Particle Physics, Central China Normal University, Wuhan 430079, China}

%%%%%%%%%%%%%%%%%%%%%%%%%%%%%%%%%%%%%%%%%%%%%%%%%%%%%%%%%%%%%%%%%%%%%%%%%%%%%

\begin{abstract}
In this paper, we study the production of isolated-photon plus a jet in $pp$ and $PbPb$ collisions, which can be used as an important probe to the jet transport property in quark gluon plasma created in heavy ion collisions. Normally, there are two types of observables associated with the production of isolated-photon plus a jet, namely, the azimuthal angular correlation and the transverse momentum imbalance. To understand both observables in the full kinematical region, we need to employ the perturbative QCD calculation, which takes into account the hard splitting of partons, together with the Sudakov resummation formalism, which resums soft gluon splittings. Furthermore, by introducing energy-loss into the system, we calculate the enhancement of the momentum imbalance distribution for $AA$ as compared to $pp$ collisions and make predictions for future unfolded experimental data. In addition, in order to extract the jet transport coefficient more precisely in our numerical calculation, we also distinguish quark jets from gluon jets, since they interact with quark gluon plasma with different strengths. This work provides a reliable theoretical tool for the calculation of the gamma-jet correlation, which can lead us to a more precise extraction of the jet transport coefficient in relativistic heavy-ion collisions.
\end{abstract}

%\pacs{24.85.+p, 12.38.Bx, 12.39.St, 12.38.Cy}
\maketitle

%%%%%%%%%%%%%%%%%%%%%%%%%%%%%%%%%%%%%%%%%%%%%%%%%%%%%%%%%%%%%%%%%%%%%%%%%%%%%

\section{Introduction}

Created in the Relativistic Heavy Ion Collider (RHIC) at Brookhaven and later at the Large Hadron Collider (LHC) at CERN, the \emph{Quark-Gluon Plasma} (QGP) is undoubtedly one of the most interesting discoveries in relativistic heavy-ion collision experiments. A lot of efforts have been devoted to unravel the mysteries of this strongly-coupled fluid, which is also related to the very early stages of the universe.

Due to multiple scatterings with QGP which induces additional gluon radiations, high energy jets traversing QGP medium can lose a significant fraction of their energy~\cite{Gyulassy:1993hr, Baier:1996kr, Baier:1996sk, Baier:1998kq, Zakharov:1996fv, Gyulassy:1999zd, Wiedemann:2000za, Arnold:2002ja, Wang:2001ifa}. In the Baier-Dokshitzer-Mueller-Peigne-Schiff-Zakharov (BDMPS-Z) jet energy loss formalism~\cite{Baier:1996kr, Baier:1996sk, Baier:1998kq, Zakharov:1996fv}, the signature of energy loss is characterized by the so-called jet transport coefficient $\hat q$~\cite{Majumder:2010qh, Qin:2015srf, Blaizot:2015lma}, which is defined as transverse momentum square transfer per unit length and reflects the density of QGP medium. In particular, early efforts in the quantitative extraction of the jet-transport coefficient from the JET collaboration were performed by calculating the nuclear modification factor ($R_{AA}$) for single hadron production using different energy-loss models, by comparing between RHIC and LHC experimental data on nucleus-nucleus ($AA$) collision and elementary hadron-hadron ($pp$) collisions \cite{Burke:2013yra}.

At the LHC, dijet transverse momentum imbalance has become an important gateway to quantitatively study the properties of quark-gluon plasma created in heavy-ion collisions. In particular, its difference between $PbPb$ and $pp$ collisions~\cite{Aad:2010bu, Chatrchyan:2011sx, Chatrchyan:2012nia} reveals that high energy jets tend to lose a significant amount of energy when traversing QGP medium created in $PbPb$ collisions~\cite{Qin:2010mn, CasalderreySolana:2010eh, Young:2011qx, He:2011pd, Lokhtin:2011qq, ColemanSmith:2012vr, Renk:2012cb, Zapp:2012ak, Ma:2013pha, Senzel:2013dta, Casalderrey-Solana:2014bpa, Ayala:2015jaa, Milhano:2015mng, Chang:2016gjp, Chen:2016cof}. 

In contrast to the dijet production, the production of isolated-photon plus a jet gives us a more clear view of jet energy loss, since only QCD jets lose energy in QGP, while the direct photon does not interact strongly with the QCD medium. As far as the quantitative study of the strongly-coupled QGP is concerned, the isolated-photon and jet correlation is considered as the golden probe of the transport properties of QGP~\cite{Wang:1991xy}. Therefore, despite its relatively small cross section, this process is complementary to the dijet productions in $pp$ and $PbPb$ collisions. In this process, the isolated-photon and jet are predominantly produced in the back-to-back region with approximately the same transverse momentum. To quantify this, normally one defines and studies the distribution of $x_{J\gamma}=\frac{p_{\perp J}}{p_{\perp \gamma}}$ with $p_{\perp J}$ and $p_{\perp \gamma}$ being the transverse momenta of the produced jet and isolated-photon, respectively. We find that the distribution of $x_{J\gamma}$ strongly peak at $1$ and the average $\langle x_{J\gamma}\rangle \simeq 1$ in $pp$ collisions. In $PbPb$ collisions, the distribution of $x_{J\gamma}$ gets smeared and $\langle x_{J\gamma}\rangle$ becomes less than $1$ due to the jet energy loss effects. Since the isolated-photon is almost undisturbed by the QGP medium, we can roughly have the direct knowledge of jet energy loss from $\langle x_{J\gamma}\rangle$. The objective of this paper is to quantitatively study the distribution of $x_{J\gamma}$ in both $pp$ and $PbPb$ collisions, which allows us to extract the value of jet transport coefficient $\hat q$ from this process. 

Recently, there have been a lot of progress in the Sudakov resummation formalism which describes the back-to-back dijet (dihadron and hadron-jet) azimuthal angular correlation~\cite{Banfi:2008qs, Mueller:2012uf, Mueller:2013wwa, Sun:2014gfa,Sun:2015doa, Mueller:2016gko, Mueller:2016xoc, Chen:2016vem}. By replacing one of the final state jet with the isolated-photon, and using proper color factors~\cite{Mueller:2013wwa} for the Sudakov factors, we can also compute the isolated-photon and jet correlation in the back-to-back region in the same formalism. When the isolated-photon and jet are far away from being back-to-back, the configuration is better described by perturbative QCD calculations of partonic matrix-elements as demonstrated in Refs.~\cite{Nagy:2001fj, Nagy:2003tz, Abazov:2004hm, Chen:2016cof}. When a pair of particles (e.g., an isolated-photon and a quark) is produced in a $2\to 2$ hard collision, their transverse momenta are balanced. In other words, the sum of their transverse momenta is zero due to momentum conservation. The momentum imbalance is mainly due to additional gluon branching in QCD, which can either occur before or after the hard collision. One can cast the gluon branchings into two categories, namely the soft branching and hard branching. Strictly speaking, the separation of these two categories is not distinct. Nevertheless, we can separate them by comparing the momentum imbalance $q_\perp$ to the jet transverse momentum $p_{\perp J} \sim p_{\perp \gamma}$(or the isolated-photon transverse momentum). If the resulting momentum imbalance $q_\perp$ is much less than $p_{\perp J}$, we can say it is a soft branching. If $q_\perp \sim p_{\perp J}$, it is considered as a hard branching. It is safe to say that the dominant contribution to the almost back-to-back configurations is the soft gluon branching. In contrast, the hard branching is much more efficient way to give the measured pair a large angle deflection. In terms of QCD calculations, the soft branchings can be approximately formulated into parton showers in Monte Carlo simulations in momentum space or taken into account by the Sudakov resummation formalism in coordinate space, which resums multiple soft gluons emissions. As to the hard branching, we still need to rely on a more complete evaluation of Feynman diagrams in terms of pQCD expansions. In fact, for each step of hard branching, one pays a price of $\alpha_s$ in QCD and goes up one more order in pQCD calculations. Therefore, as far as the momentum imbalance and the angular correlation are concerned, we need to employ both the Sudakov resummation formalism and pQCD calculations in order to understand them in full phase space~\cite{Chen:2016cof}. 

At RHIC, using leading hadrons as surrogates of jets, the study of isolated-photon and jet correlations is pioneered by some early work on photon-hadron correlations~\cite{Wang:1996yh, Wang:1996pe, Renk:2006qg, Zhang:2009rn, Qin:2009bk, Adare:2009vd,  Abelev:2009gu, Chen:2017zte}. In the era of the LHC, most of the previous theoretical studies\cite{Qin:2012gp, Wang:2013cia, Ma:2013bia, Luo:2018pto, Kang:2017xnc} of isolated-photon (or vector boson) and jet correlations are based on Monte Carlo (MC) simulations of parton showers (PS), since the dominant contribution comes from the back-to-back photon-jet configurations. In the back-to-back region, the main source of transverse momentum imbalance is due to multiple soft branchings as mentioned earlier. However, soft radiations may not be a sufficient way to get large momentum imbalances. With one step of hard branching, we can easily obtain a significant momentum imbalance between the isolated-photon and measured jet. In addition, parton showers ignore quantum interference between scattering amplitudes. It is reasonable to expect that, after a large number of soft branchings, the quantum interferences of matrix-elements is negligible and the factorisation of soft gluons from the lowest order hard part is safe, but the parton showers on top of the lowest order calculations often become insufficient for a lot of observables at the LHC (e.g., the $x_{J \gamma}$ distribution when it is away from $1$).

On the other hand, one can use the MC matrix-element event generator such as ``JETPHOX''~\cite{Catani:2002ny, Belghobsi:2009hx} to compute the so-called next-to-leading (NLO) matrix element (up to $2\to 3$ processes) for photon productions\footnote{As far as the isolated-photon and jet correlation is concerned, this is only considered as leading order(LO) in pQCD calculation, since the tree level $2\to 2$ process gives trivial results for the correlation. In the Sudakov resummation formalism, all the $2\to n$ processes with $n\geq 2$ have been approximately taken into account in the soft gluon limit.}, as first studied in Ref.~\cite{Dai:2012am}. However, this approach only has a few particles in the final state and it does not take into account the parton shower effect, which is indispensable to describe the back-to-back configurations. 

These two approaches are complementary to each other in modern high energy physics research. The efforts of merging higher-order matrix-element event generator and parton showers have attracted a lot of attention in the development of MC generators recently (e.g., POWHEG~\cite{Alioli:2010xa}). In this work, without using MC generators, we adopt a physically equivalent theoretical framework, which also takes into account both higher order matrix-elements in pQCD and parton showers (in terms of Sudakov resummations). For the calculation of photon-jet correlations considered in this paper, by computing $2\to 3$ matrix elements and employing Sudakov resummation formalism, we can achieve the same NLO accuracy as the modern MC generators. Furthermore, for dijet productions, as shown in our last paper~ \cite{Chen:2016cof}, we have reached NNLO accuracy in terms of dijet asymmetries, which outperforms the NLO MC generators.

In this paper, following the formalism developed in Ref.~\cite{Chen:2016cof}, we present the a systematic investigation of photon-jet correlations based on pQCD calculations supplemented with the Sudakov resummation in the soft gluon region 
as the baseline calculation in $pp$ collisions. The result for angular correlations is in agreement with both ATLAS and CMS data, while the result for the $x_{J\gamma}$ distribution clearly differs from experimental data. The latter case is due to the fact that the $x_{J\gamma}$ distribution is sensitive to detector effects which have not yet been removed from the current experimental data. Interestingly, we discover that our results for the $x_{J\gamma}$ distribution coincide with the Monte Carlo results\cite{Klasen:2017dsy} obtained by using POWHEG+PYTHIA. This is expected, since the basic ingredients of POWHEG and our Sudakov resummation improved pQCD framework are the same, as far as the $x_{J\gamma}$ distribution is concerned. To compare with the ATLAS data, we employ a smearing function within the given range of the ATLAS correction to mimic the detector response on top of our $pp$ calculation, and we found excellent agreement with the ATLAS data. In the end, when the energy loss effect based on the BDMPS formalism is implemented, we find the transport coefficient $\hat{q}_0 \sim 2 \, -\, 8\textrm{GeV}^2/\textrm{fm}$ for $PbPb$ collisions at the LHC at $5.02$TeV.

%%%%%%%%%%%%%%%%%%%%%%%%%%%%%%%%%%%%%%%%%%%%%%%%%%%%%%%%%%%%%%%%%%%%%%%%%%%%%

\section{Isolated-photon and jet azimuthal angular correlation} 

In this study, we will investigate both the azimuthal angular decorrelation and momentum imbalance of isolated-photon tagged jets, and study energy-loss effect of final state jets.
By extending our previous formalism of the Sudakov resummation improved perturbative QCD approach, we can provide a much more quantitative analysis on the gamma-jet angular correlation, and make relevant prediction on the unfolded momentum imbalance distribution.

%-------------------------------------------------------------%
In this section, we will investigate the azimuthal angular decorrelation between the triggered photon and the associate jet, namely the $\Delta\phi=|\phi_\gamma-\phi_J|$ distribution. This is also important in understanding the transverse momentum broadening effect of the QGP medium. We begin by writing the differential cross-section of the angular distribution in the Sudakov resummation formalism as follows
\begin{align}
\frac{d\sigma}{d\Delta\phi}&=\sum_{a,b,c,d}\int p_{\perp\gamma}dp_{\perp\gamma}\int p_{\perp\! J}dp_{\perp J}\int dy_{\gamma}\int dy_J\int db \notag \\
&~~~~\times x_af_a(x_a,\mu_b)x_bf_b(x_b,\mu_b)\frac{1}{\pi}\frac{d\sigma_{ab\rightarrow cd}}{d\hat{t}} b~J_0(|\vec{q}_\perp|b)e^{-S(Q,b)} ,
\end{align}
where $J_0$ is the Bessel function of the first kind, $q_\perp$ is the transverse momentum imbalance between the photon and the jet $\vec{q}_\perp\equiv\vec{p}_{\perp\gamma}+\vec{p}_{\perp J}$, which takes into account both initial and final transverse momentum kicks from vacuum Sudakov radiations and medium gluon radiations.
Here we define $x_{a,b}=max(p_{\perp\gamma},p_{\perp J})(e^{\pm y_\gamma}+e^{\pm y_J})/\sqrt{s_{NN}}$ as the momentum fraction of the incoming parton $a,b$ from the parent nucleon. $f_{a,b}(x,\mu_b)$ are the parton distribution functions (PDFs) of the incoming parton, and $d\sigma/d\hat{t}$ is the leading order partonic cross-section. In this study, the cross-section comes from the sum of $gq\rightarrow \gamma q$ and $q\bar{q}\rightarrow \gamma g$ sub-processes. Since the Sudakov soft gluons can be radiated from both the incoming and outgoing hard partons, it is impossible to distinguish individual contributions to the momentum imbalance $\vec{q}_\perp$. An attempt to approximate and reconstruct the incoming partonic information is to calculate the momentum fraction $x_{a,b}$ and the Mandelstam variables $s,t,u$ using the maximum outgoing particle $p_{\perp}$ as above.

The vacuum Sudakov factor $S_{pp}(Q,b)$ is defined as
\begin{equation}
S_{pp}(Q,b)=S_P(Q,b)+S_{NP}(Q,b)
\end{equation}
where the perturbative $S_P$ Sudakov factor depends on the incoming parton flavour and outgoing jet cone size.
The perturbative Sudakov factors can be written as~\cite{Mueller:2013wwa, Sun:2014gfa,Sun:2015doa}
\begin{equation}
S_P(Q,b)=\sum_{q,g}\int_{\mu_b^2}^{Q^2}\frac{d\mu^2}{\mu^2}\left[A\ln\frac{Q^2}{\mu^2}+B+D\ln\frac{1}{R^2}\right]
\end{equation}
At the next-to-leading-log (NLL) accuracy, the coefficients can be expressed as $A=A_1\frac{\alpha_s}{2\pi}+A_2(\frac{\alpha_s}{2\pi})^2$, $B=B_1\frac{\alpha_s}{2\pi}$ and $D=D_1\frac{\alpha_s}{2\pi}$, with the value of individual terms given by the following table, where both $A$ and $B$ terms are summed over the corresponding incoming parton flavours.
\begin{center}
\vspace{-5pt}
\begin{tabular}{c||c|c|c|c}
&	$A_1$	&	$A_2$			&	$B_1$				&	$D_1$	\\\hline
quark	&	$C_F$	&	$K\cdot C_F$	&	$-\frac{3}{2}C_F$	&	$C_F$	\\
gluon	&	$C_A$	&	$K\cdot C_A$	&	$-2\beta C_A$		&	$C_A$
\end{tabular}
\end{center}
Here $C_A$ and $C_F$ are the gluon and quark Casimir factor, respectively. $\beta=\frac{11}{12}-\frac{N_f}{18}$, and $K=(\frac{67}{18}-\frac{\pi^2}{6})C_A-\frac{10}{9}N_fT_R$. $R^2=\Delta\eta^2+\Delta\phi^2$ represents the jet cone-size, which is set to match the experimental setup. The implementation of the non-perturbative Sudakov factor $S_{NP}(Q,b)$ follows the prescription given in Refs~\cite{Su:2014wpa, Prokudin:2015ysa}. In the Sudakov resummation formalism, following the usual $b^{\ast}$ prescription, the factorization scale is set to be $\mu_b\equiv\frac{c_0}{b_\perp}\sqrt{1+b_\perp^2/b_{max}^2}$, where $c_0=2e^{-\gamma_E}$ and $b_{max}=0.5\textrm{GeV}^{-1}$ which is chosen to separate the perturbative region from the non-perturbative region. The strong coupling runs with the hard scale $Q^2=x_ax_bs$. As suggested in Ref.~\cite{Qiu:2000ga}, the Sudakov effect is dominated by perturbative contributions and insensitive to the choices of non-perturbative parts, when the hard scale $Q$ is sufficiently large. We have also confirmed this in our numerical calculation. 

For $AA$ collisions, by adding additional transverse momentum broadening due to the interactions between QGP and outgoing jets as suggested in BDMPS formalism, one can simply adopt the following form of the Sudakov factor,
\begin{equation}
S_{AA}(Q,b)=S_{pp}+\hat{q}_R L\frac{b^2}{4}
\end{equation}
where $\hat{q}_R=\hat{q}_q$ or $\hat{q}_g$ corresponds to quark and gluon jets transverse momentum broadening, respectively. One can relate them as $\hat{q}_g=\frac{C_A}{C_F}\hat{q}_q$. The above expression separates the medium broadening effect from the vacuum Sudakov effect. This is due to the fact that both effects have well-separated regions in their phase space integral which contribute differently to the transverse momentum broadening effects~\cite{Mueller:2016gko, Mueller:2016xoc}.

As to direct photon productions in the pQCD framework~\cite{Owens:1986mp}, we use the results computed from $2\to 3$ matrix elements in Refs.~\cite{Baer:1989xj, Baer:1990ra} to compute the productions of the isolated-photon plus a jet $$p+p\to \gamma (\phi_\gamma, p_{\perp \gamma}) +J(\phi_J, p_{\perp, J}) +X.$$ The correlation is normalized by the LO ($2\to 2$) inclusive isolated-photon cross section as common practices. In pQCD framework, it is the third unobserved particle $X$ which provides the momentum imbalance between the isolated-photon and the measured jet. The corresponding isolation cut around the isolated-photon has been applied to the calculation to match the data selection of ATLAS and CMS experiments. In spite of strong dependence on the choice of scales $\mu$ in pQCD calculations, we find that our results give nice agreement with experimental data when we set $\mu =Q$, which is the most obvious choice. When $\Delta\phi=|\phi_\gamma-\phi_J|$ is sufficiently away from $\pi$, the momentum of $X$ is sufficiently large to ensure the calculation is indeed perturbative. However, when $\Delta\phi \sim \pi$ which implies the isolated-photon and the measured jet are back-to-back, the third unobserved particle $X$ becomes soft which introduces large Sudakov type logarithms. Therefore, the Sudakov resummation must be employed to regain the predictive power in QCD. We would like to emphasize that the normalization of pQCD calculation is carried out in the sense of perturbative expansion following the convention in Refs.~\cite{Nagy:2001fj, Nagy:2003tz, Abazov:2004hm, Chen:2016cof}. Here we use the LO $2\to 3$ results for angular correlation as the numerator, and use the LO $2\to 2$ results for inclusive total cross section as the denominator to obtain 
the `normalized' pQCD $\Delta \phi$ distribution. One should keep in mind that $2\to 2$ process is discarded as far as angular correlations are concerned. 

\begin{figure}[tbp]
\begin{center}
\includegraphics[width=0.32\linewidth]{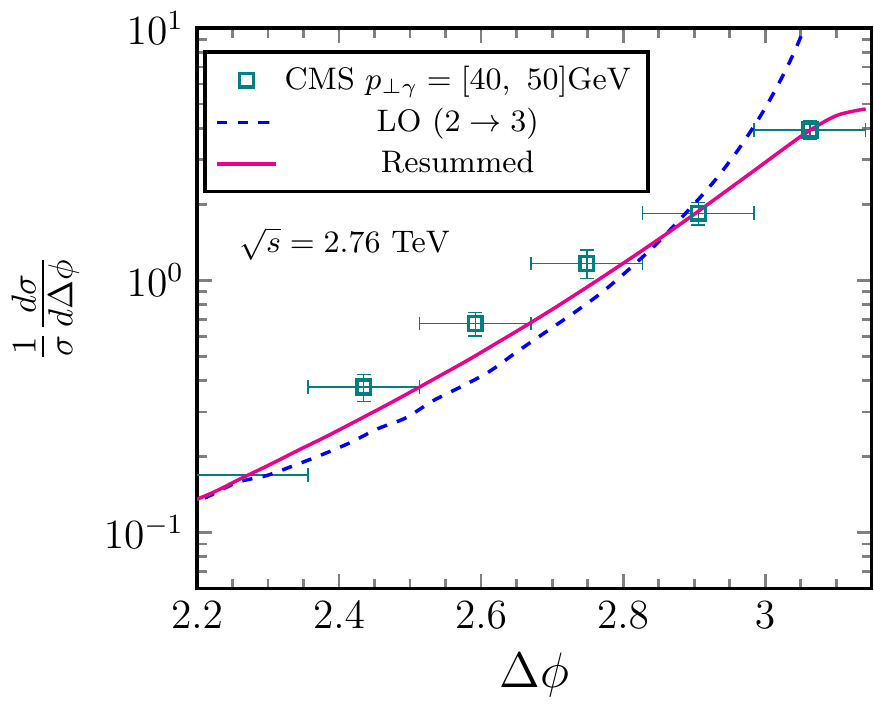}
\includegraphics[width=0.32\linewidth]{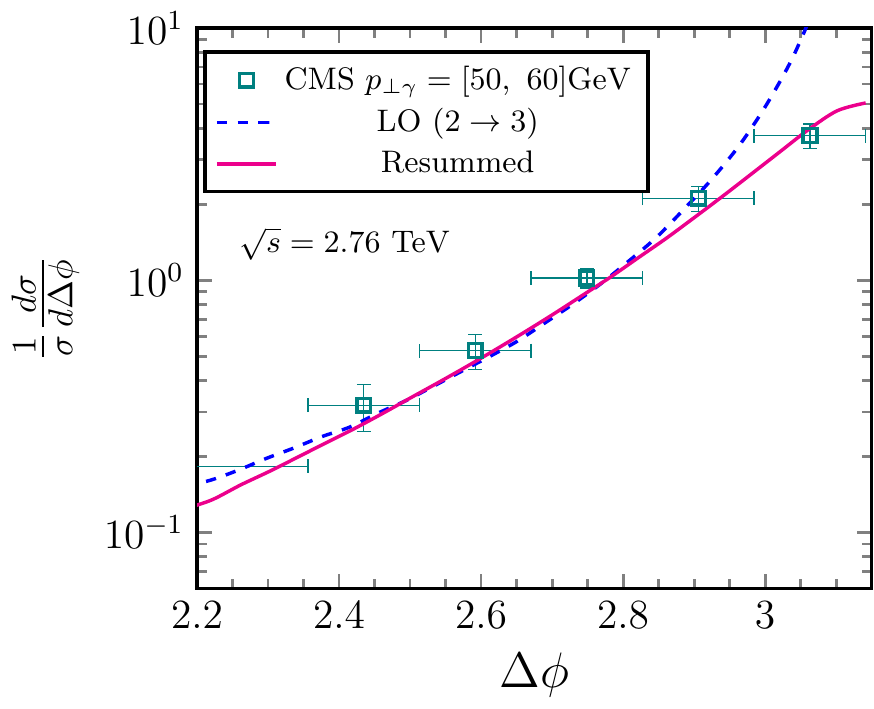}
\includegraphics[width=0.32\linewidth]{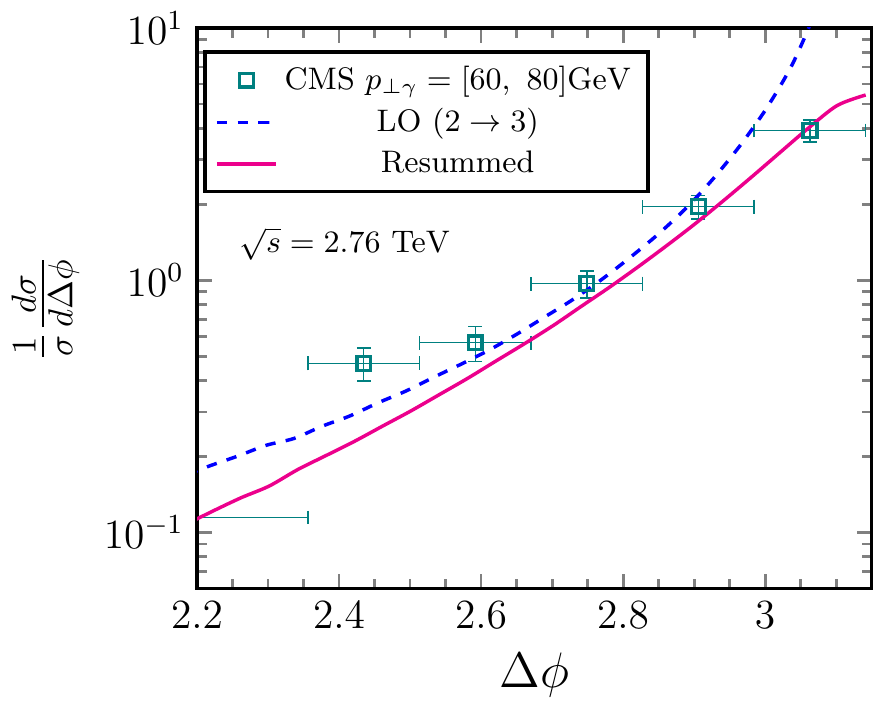}
\includegraphics[width=0.32\linewidth]{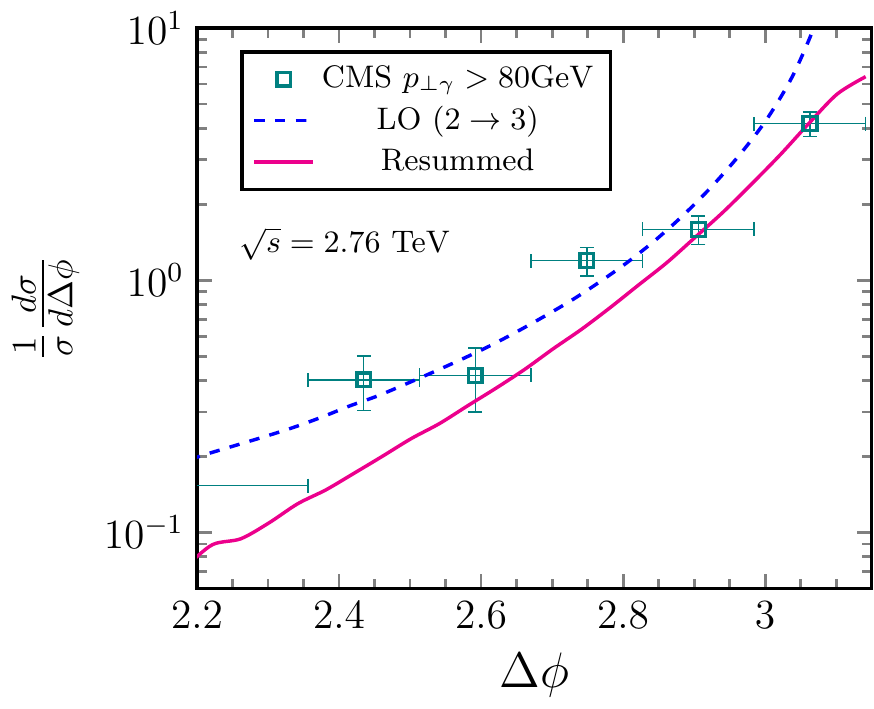}
\includegraphics[width=0.32\linewidth]{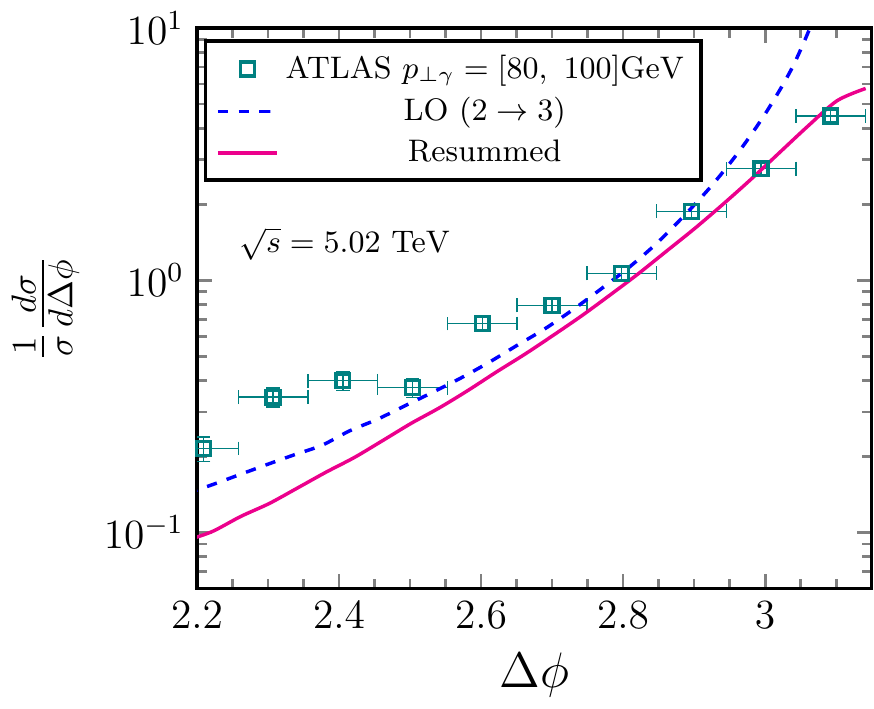}
\includegraphics[width=0.32\linewidth]{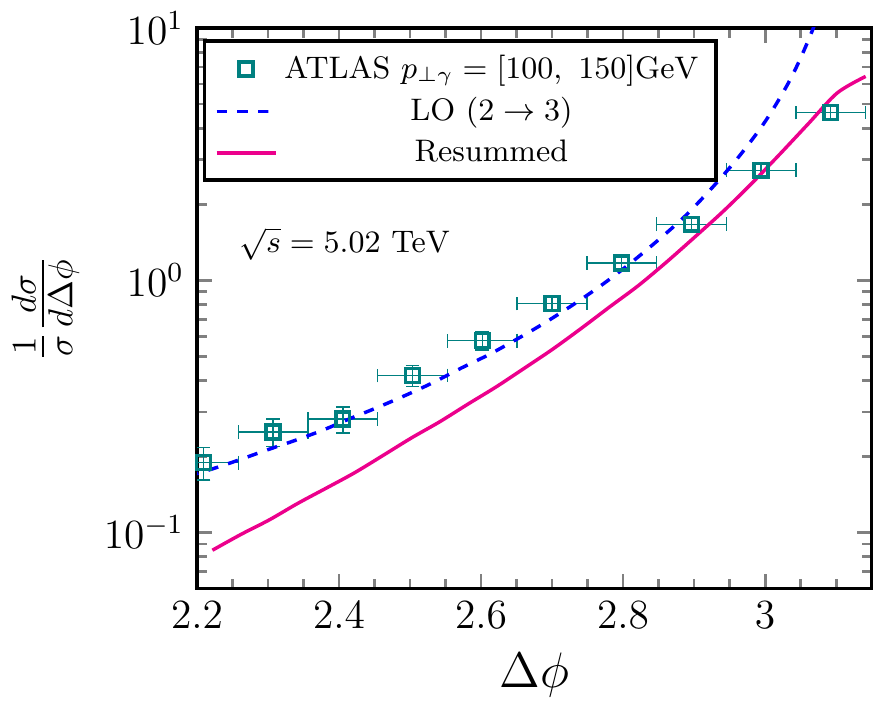}
\end{center}
\caption[*]{The normalized $\Delta \phi$ distributions in six regions of $p_{\perp \gamma}$ at $2.76$~TeV measured by CMS~\cite{CMS:2013oua} with $ p_{\perp J}  >30~\textrm{GeV}$, $R=0.3$, rapidity cut $|y|<1.6$ and at $5.02$~TeV measured by ATLAS~\cite{ATLAS:2016tor} with $ p_{\perp J}  >30~\textrm{GeV}$, $R=0.4$ and rapidity cut $|y|<2.1$.}
\label{phi}
\end{figure}

By calculating the photon-jet angular correlations in both the pQCD framework and Sudakov resummation formalism, we can plot the normalized $\Delta\phi$ distribution and compare them with the ATLAS and CMS data~\cite{Chatrchyan:2012gt, CMS:2013oua, ATLAS:2016tor, Sirunyan:2017qhf} as shown in Fig.~\ref{phi}. This normalized distribution is known to be insensitive to hadronization corrections and the underlying event~\cite{Abazov:2004hm}. In addition, since we also integrate over a large range of transverse momentum, the $\Delta\phi$ distribution seems to be insensitive to the detector effects as well. Let us first pay attention to the LO pQCD calculation for the angular correlation ($2\to 3$ processes). In the large deflection angle region where $\Delta\phi$ is away from $\pi$, the pQCD results describe all the experimental data well. As expected, the pQCD calculation clearly diverges in the region near $\Delta\phi=\pi$, while the Sudakov resummation converges to a finite value. This is mainly due to the fact that the Sudakov type logarithms $\alpha_s \ln^2 \frac{p_{\perp \gamma}^2}{q_\perp^2}$ become very large in the back-to-back region, which makes the naive perturbative expansions inadequate to describe this region. As a result, we find that the Sudakov resummation results describe the back-to-back region very well. 

We have also examined the angular correlations with the additional medium transverse momentum broadening due to QGP medium. We also found almost identical curves for the results in $pp$ and $PbPb$ collisions even with $\hat{q}_q L \sim 20 \textrm{GeV}^2$. This is consistent with all the experimental results at the LHC~\cite{Chatrchyan:2012gt, CMS:2013oua, ATLAS:2016tor, Sirunyan:2017qhf}, in which essentially no difference is observed in the back-to-back region between the angular correlation data collected in $pp$ and $PbPb$ collisions. As first quantitatively pointed out in Ref.~\cite{Mueller:2016gko}, this is due to the fact that the vacuum Sudakov effects in $pp$ collisions are so overwhelming in the LHC kinematics that the QGP medium transverse momentum broadening effects are simply negligible. Nevertheless, at lower collision energies such as at RHIC kinematics, the magnitudes of both vacuum and medium effects are of similar order and thus, a significant medium broadening could be observed. We also note here the jet energy-loss effects do not change the azimuthal angular distribution much. In contrast, the medium energy loss effect plays an important role in the transverse momentum imbalance calculation as shown below.

%-------------------------------------------------------------%

\section{Photon-Jet Momentum Imbalance}

Similar to the findings in the dijet productions, the momentum imbalance measurement, which is known as the dijet asymmetry, is prone to a significant modification due to QGP medium effects. Let us now consider the gamma-jet momentum imbalance distribution, which is characterized by the variable $x_{J\gamma}\equiv{p_{\perp J}}/{p_{\perp\gamma}}$. In our previous study of dijet asymmetries~\cite{Chen:2016cof}, we have shown that the dijet asymmetry ratio in perturbative calculations obeys a distribution bound depending on the number of final state particles produced. The gamma-jet momentum imbalance does not follow such bound since the photon and jet transverse momenta are not ordered.

Note that Sudakov resummation alone does not give a good description of the $x_{J\gamma}$ distribution, since a large region of the distribution (away from the region $x_{J\gamma} \sim 1$) is dominated by the perturbative QCD calculation.  Thus both pQCD and Sudakov formalism should be used. By implementing the resummation improved pQCD approach developed in Ref.~\cite{Chen:2016cof}, we can compute the gamma-jet momentum imbalance distributions as follows
\begin{eqnarray}
%  \left.\frac{1}{\sigma} \frac{d\sigma}{dx_J}\right|_{\textrm{improved}} = \left.\frac{1}{\sigma_{\textrm{pQCD}}} \frac{d\sigma_{\textrm{pQCD}}}{dx_J}\right|_{\Delta \phi <\phi_{\textrm{m}}}
% +\left.\frac{1}{\sigma_{\textrm{Sudakov}}} \frac{d\sigma_{\textrm{Sudakov}}}{dx_J}\right|_{\phi_{\textrm{m} < \Delta \phi <\pi}}, \label{improved}
 \left.\frac{1}{\sigma} \frac{d\sigma}{dx_J}\right|_{\textrm{improved}} = \left.\frac{1}{\sigma_{\textrm{pQCD}}} \frac{d\sigma_{\textrm{pQCD}}}{dx_J}\right|_{\rm cuts}
+\left.\frac{1}{\sigma_{\textrm{Sudakov}}} \frac{d\sigma_{\textrm{Sudakov}}}{dx_J}\right|_{\rm cuts}, \label{improved}
\end{eqnarray}
where the momentum imbalance in the so-called `improved' approach is computed as the overlay of the pQCD results and Sudakov resummed results, which are defined as
\begin{eqnarray}
%\frac{d\sigma}{d\Delta \phi} &=&\int \frac{d\sigma}{d^2p_{\perp 1}d^2p_{\perp 2}} d^2p_{\perp 1} d^2p_{\perp 2} \delta (\Delta \phi -\phi_1 +\phi_2), \,\\% (2\pi)p_{\perp 1}p_{\perp 2} dp_{\perp 1} dp_{\perp 2}, \\
\frac{d\sigma_\textrm{pQCD/Sudakov}}{dx_{J\gamma}} =\int d^2p_{\perp \gamma} d^2p_{\perp J} \delta \left(x_{J\gamma} -\frac{p_{\perp J}}{p_{\perp \gamma}}\right) \left.\frac{d\sigma_\textrm{pQCD/Sudakov}}{d^2p_{\perp \gamma}d^2p_{\perp J}} \right|_{\rm cuts} . \label{xjg}
\end{eqnarray}
To match the Sudakov resummation to the pQCD calculation, we first cut the phase space into two regions, namely, the $\Delta \phi < \phi_{\rm m}$ part and $\phi_{\rm m}< \Delta\phi < \pi$ part. When $\Delta \phi$ is smaller than the matching point, where $\phi_{\rm m} = 2.9$, we apply the perturbative QCD calculation. In the $\phi_{\rm m}< \Delta\phi < \pi$ region, Sudakov resummation gives precise description on the angular distribution. Nevertheless, to compute the $x_{J\gamma}$ distribution, we need to further impose a cut on $q_\perp <q_{\textrm{cut}}$ so that we can make sure that the prerequisite $q_{\perp}^2 \ll P_\perp^2$ is satisfied for all the events the $\phi_{\rm m}< \Delta\phi < \pi$ region calculated in terms of the Sudakov resummation. Note that if $q_{\perp}^2 \sim P_\perp^2$, the Sudakov resummation is no longer applicable. Therefore we further separate the phase space into large $q_\perp$ part and small $q_\perp$ part. The Sudakov resummation is turned on only at small $q_\perp$ region. When $q_\perp$ is larger than the semi-hard scale $q_{\textrm{cut}}$, which is chosen as $p_{\perp \gamma}/4$, we employ the pQCD calculation again. We would like to emphasize here that the hard scale can not be very small. Otherwise the double logarithm $\ln^2 \frac{Q^2}{q_\perp^2}$ will break the perturbative expansion. As long as the scale is reasonably large enough, the finally result will not strongly depend on the choice of $q_{\textrm{cut}}$. %since the dominate contribution comes from small $q_\perp$ part.

\begin{figure}[tbp]
\begin{center}
\includegraphics[width=0.32\linewidth]{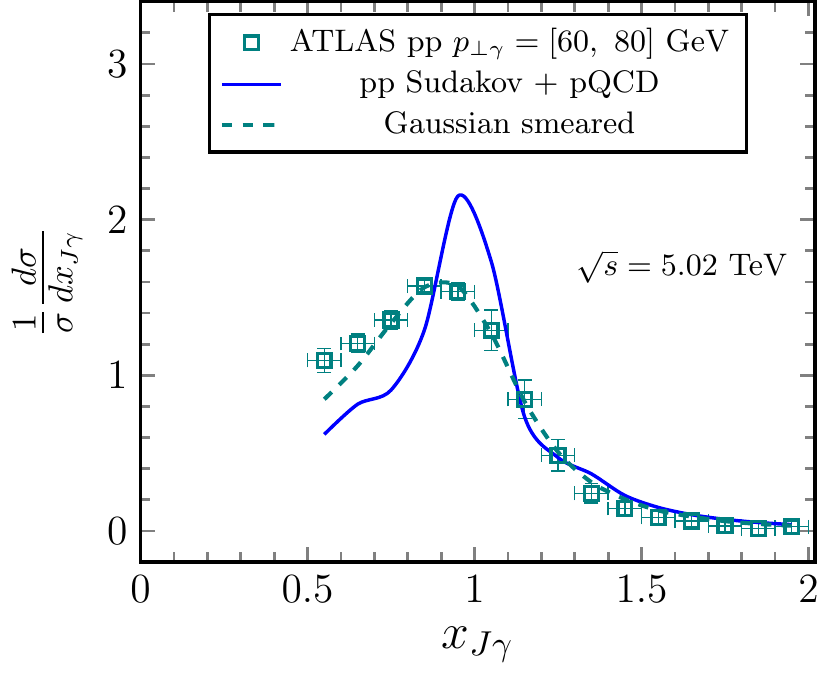}
\includegraphics[width=0.32\linewidth]{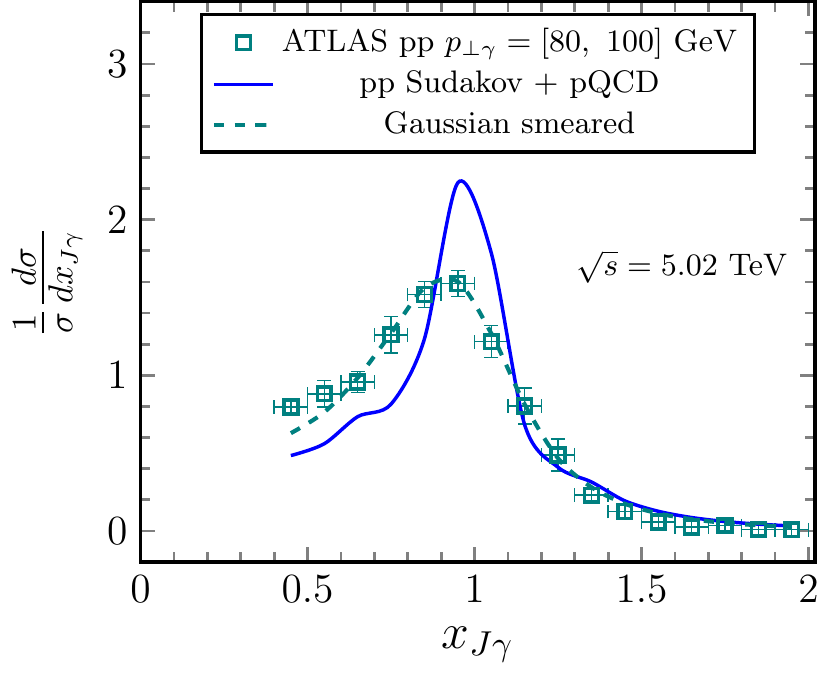}
\includegraphics[width=0.32\linewidth]{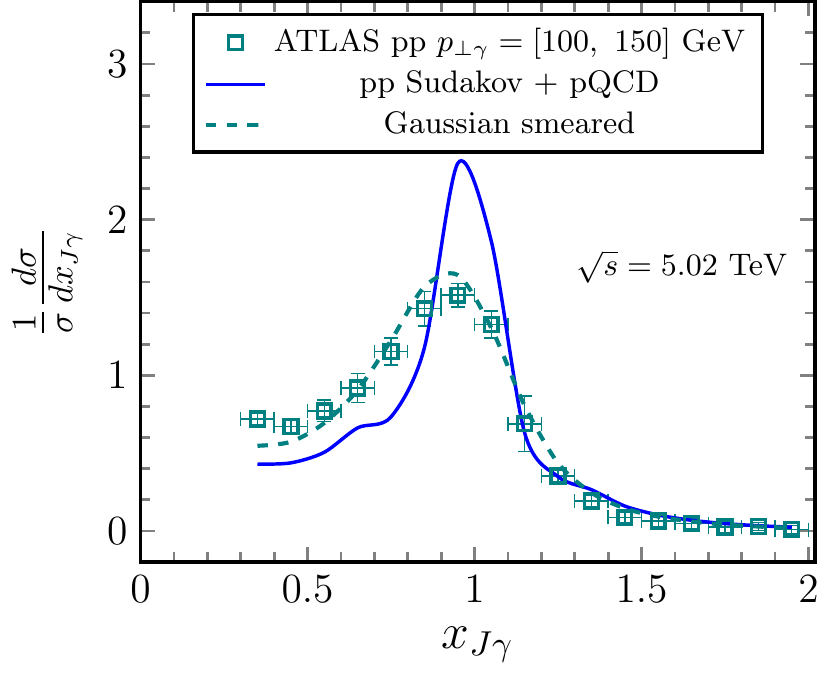}
\end{center}
\caption[*]{Normalized $x_{J\gamma}$ distributions computed for $pp$ collisions at $5.02$~TeV with $R=0.4$, $p_{\perp J}> 30\textrm{GeV}$ and jet rapidity cut $|y|<2.1$ compared with the ATLAS data taken from Ref.~\cite{ATLAS:2016tor}. The cuts on the isolated photon is set to be the same as the ATLAS data selection. 
In order to compare with the ATLAS data with the same bin selection, we have removed all events with $x_{J\gamma}$ smaller than the minimal $x_{J\gamma}$ bin chosen in Ref.~\cite{ATLAS:2016tor}.
}
\label{atlasxjg1}
\end{figure}

Second, if we compare our results in $pp$ collisions with the ATLAS experimental data directly as shown in Fig.~\ref{atlasxjg1}, we find that our calculation is a bit far away from the data, especially for the high $p_{\perp \gamma}$ bins. Nevertheless, this never means that our QCD calculation is inadequate to describe the momentum imbalance data. As a matter of fact, we believe that the discrepancy is mainly due to the fact that all the experimental data in this type of measurement has not been fully corrected. Unlike the angular correlation measurement, the momentum imbalance measurement requires precise knowledge of jet energy which suffers a lot from effects of detector response and underlying events. It is well-known that the so-called unfolding procedure has to be applied to the data analysis to obtain the fully corrected data. A clear indication would be that for $pp$ collisions without medium effect, the $x_{J\gamma}$ distribution should peak near $1$ with the dominant effect coming from back-to-back configuration, where both photon and jet carry equal magnitude of $p_T$ according to the transverse momentum conservation. However, without the process of unfolding, the detector picks up both $p_T$ with some uncertainty, this causes bin migration to small $x_{J\gamma}$ and results in a shift of the overall distribution. There is a very interesting example in dijet productions which illustrates how important these unfolding corrections are. As demonstrated in Ref.~\cite{unfold, Perepelitsa:2016zbe}, ATLAS collaboration has successfully unfolded their data in the dijet asymmetry measurement and found that the uncorrected data underestimates the asymmetries in the back-to-back region by a large extent.  We also found that very good agreement between our theoretical calculation~\cite{Chen:2016cof} and unfolded dijet asymmetry data in our latest work. The situation in the isolated-photon and jet momentum imbalance measurement is quite similar, and the same pattern of change to the data is expected to occur. In the end, we believe that the agreement between our results given in terms of solid curves and experimental data will be significantly improved, once the unfolding procedure is carried out\footnote{According to the private communication with colleagues from the ATLAS collaboration, the isolated-photon and jet imbalance data will also be unfolded for bothe $pp$ and $AA$ in the near future.}. 

Therefore, based on the current available data, we will proceed as follows. On one hand, we provide our original results for $pp$ and $PbPb$ collisions, which should only be compared with the fully corrected data in the future. This serves as our predictions based on QCD calculations and modelling of QGP. On the other hand, as a temporary solution, we convolute our results for $pp$ collisions with a one-dimensional Gaussian smearing function as follows
\begin{equation}
\frac{d\sigma_\textrm{smeared}}{dp_{\perp J}}=\int \frac{dE}{\sqrt{2\pi}\Delta} e^{-\frac{(E-\bar E)^2}{2\Delta^2}} \left.\frac{d \sigma}{dp^\prime_{\perp J}}\right|_{p^\prime_{\perp J}=p_{\perp J}+E}.  \label{s1}
\end{equation}
$\bar E = 0.03 p^\prime_{\perp J}$ and $\Delta = 0.13p^\prime_{\perp J}$.
Another equivalent way to parametrize the smearing is
\begin{equation}
\frac{d\sigma_\textrm{smeared}}{dp_{\perp J}}=\int \frac{dr}{\sqrt{2\pi}\sigma} e^{-\frac{(r-\bar r)^2}{2\sigma^2}} 
\int dp^\prime_{\perp J} \delta(p_J-rp^\prime_{\perp J})
\frac{d \sigma}{dp^\prime_{\perp J}} 
= \int \frac{dr}{\sqrt{2\pi}\sigma} e^{-\frac{(r-\bar r)^2}{2\sigma^2}} \left. \frac{1}{r} \frac{d \sigma}{dp^\prime_{\perp J}} \right|_{p^\prime_{\perp J}=p_{\perp J}/r} .  \label{s2}
\end{equation}
with $\bar r = 0.97$ and $\sigma = 0.13$. This corresponds to a value of $3\%$ for the jet energy scale and a value of $0.13$ for the jet energy resolution, which are fully in agreement with the values used in the ATLAS unfolding analysis\cite{ATLAS:2012cna, Aad:2012ag}.

% \begin{figure}[tbp]
% \begin{center}
% \includegraphics[width=0.32\linewidth]{generated-figures/cms-xg-40-smeared-gaussian}
% \includegraphics[width=0.32\linewidth]{generated-figures/cms-xg-50-smeared-gaussian}
% \includegraphics[width=0.32\linewidth]{generated-figures/cms-xg-60-smeared-gaussian}
% \includegraphics[width=0.32\linewidth]{generated-figures/cms-xg-80-smeared-gaussian}
% \includegraphics[width=0.32\linewidth]{generated-figures/cms-xg-40-5-smeared-gaussian}
% \includegraphics[width=0.32\linewidth]{generated-figures/cms-xg-50-5-smeared-gaussian}
% \includegraphics[width=0.32\linewidth]{generated-figures/cms-xg-60-5-smeared-gaussian}
% \includegraphics[width=0.32\linewidth]{generated-figures/cms-xg-80-5-smeared-gaussian}
% \includegraphics[width=0.32\linewidth]{generated-figures/cms-xg-100-5-smeared-gaussian}
% \end{center}
% \caption[*]{Normalized $x_{J\gamma}$ distributions computed for $pp$ collisions at $2.76$~TeV and $5.02$~TeV with $R=0.3$ and jet rapidity cut $|y|<1.6$ compared with the CMS data taken from Ref.~\cite{CMS:2013oua, Sirunyan:2017qhf}.}
% \label{cmsxjg1}
% \end{figure}

\begin{figure}[tbp]
\begin{center}
\includegraphics[width=0.32\linewidth]{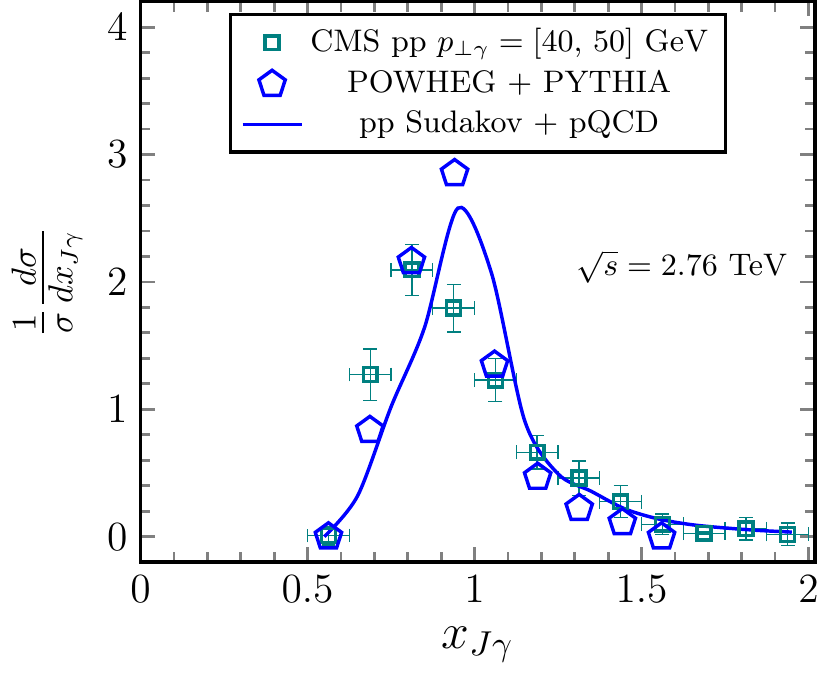}
\includegraphics[width=0.32\linewidth]{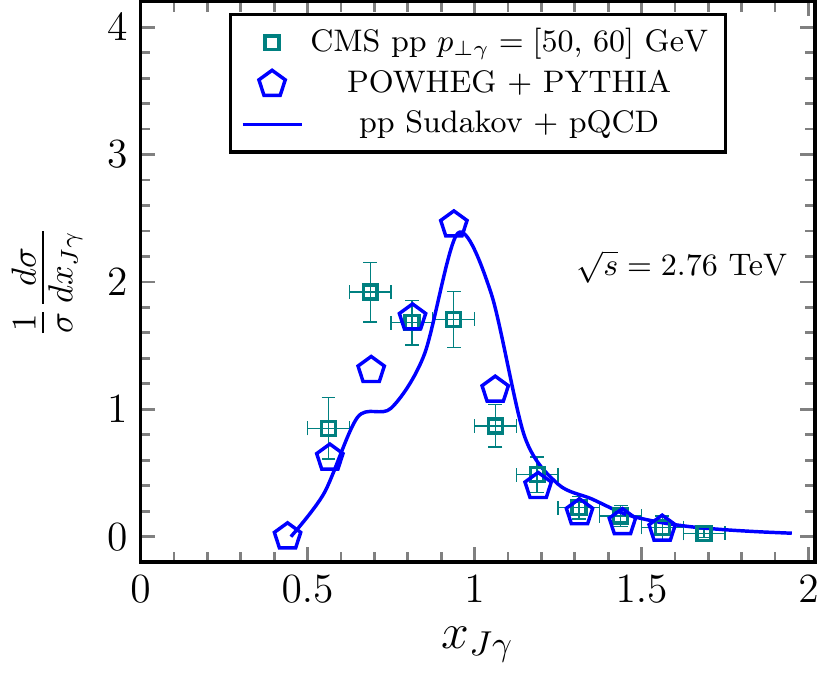} \\
\includegraphics[width=0.32\linewidth]{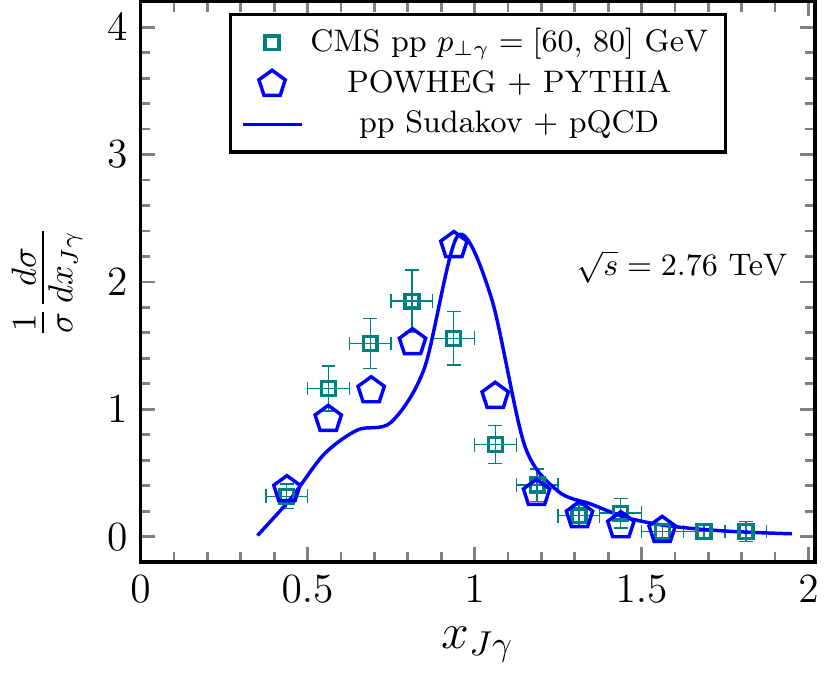}
\includegraphics[width=0.32\linewidth]{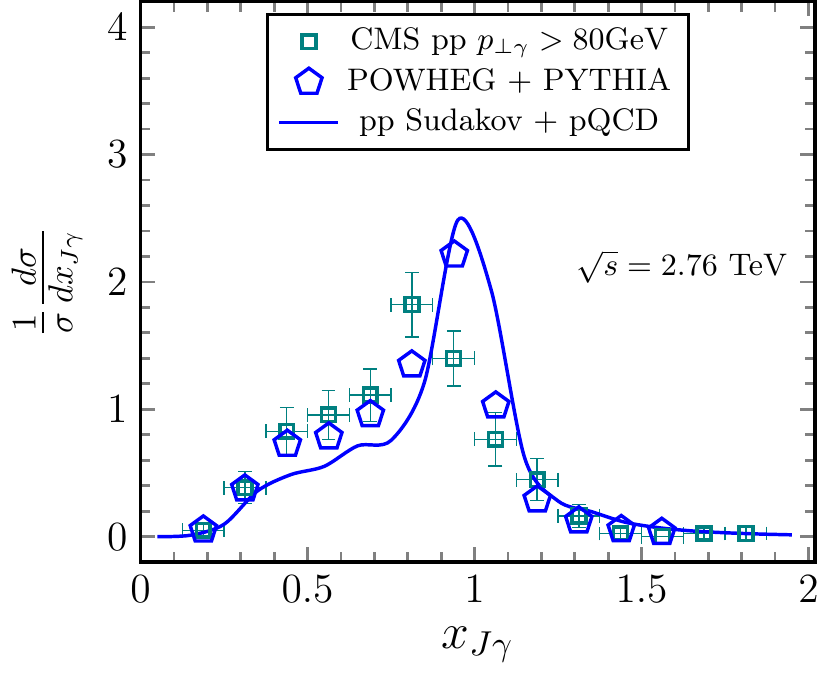}
\end{center}
\caption[*]{Normalized $x_{J\gamma}$ distributions computed for $pp$ collisions at $2.76$~TeV compared with the Monte Carlo results obtained by using POWHEG $+$ PYTHIA\cite{Klasen:2017dsy} and the CMS data taken from Ref.~\cite{CMS:2013oua}. Throughout this paper, pentagon symbols are used to represent Monte Carlo simulation results, while the square and circle symbols are utilized as labels for the experimental data measured in $pp$ and $AA$ collisions, respectively.}
\label{cmsxjg12}
\end{figure}

Using the above smearing function, we are able to describe the measured $pp$ data quite well as shown Fig.~\ref{atlasxjg1}. We see that the gamma-jet momentum imbalance distribution has a very clear peak at $x_{J\gamma}\sim 1$ and an exponential tail in the $x_{J\gamma}>1$ region. Interestingly, a shoulder starts to develop in the vicinity of the $x_{J\gamma}\sim 0.5$ region mostly due to pQCD contributions. For example, it is likely that a direct photon can be produced in an event accompanied by two qualified jets which pass the analysis selection criteria in high energy collisions. The shoulder is generated since both jets are taken into account when $x_{J\gamma}$ is measured. Due to the presence of $30$GeV $p_\perp$ cut on jets, it only becomes visible when the transverse momentum of direct photon becomes large enough. 

Furthermore, as shown in Fig.~\ref{cmsxjg12}, we compared our results with the Monte Carlo results obtained by using the combination of POWHEG+PYTHIA \cite{Klasen:2017dsy}, and found excellent quantitative agreement. This indicate that our calculation for $pp$ collisions without any smearing is equivalent to a NLO Monte Carlo generator. In contrast, we found that our $pp$ results can not reproduce the CMS $pp$ data\cite{CMS:2013oua} unless we use very large shift ($\sim 11 \%$) for the jet energy scale in Eqs.~(\ref{s1},\ref{s2}) , which moves the curves to the left significantly. Similar pattern has also been found for the comparison with the data from Refs.~\cite{CMS:2013oua, Sirunyan:2017qhf}. Therefore, we leave the comparison with CMS data for a future study.

%%%%%%%%%%%%%%%%%%%%%%%%%%%%%%%%%%%%%%%%%%%%%%%%%%%%%%%%%%%%%%%%%%%%%%%%%%%%%

\section{Photon and jet transverse momentum imbalance in $PbPb$ collisions}

\begin{figure}[tbp]
\begin{center}
\includegraphics[width=0.32\linewidth]{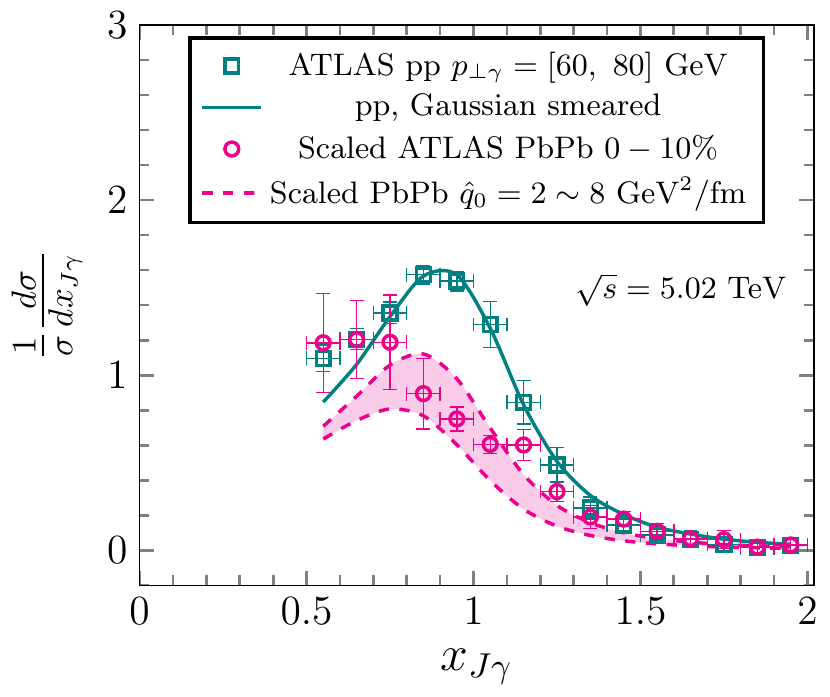}
\includegraphics[width=0.32\linewidth]{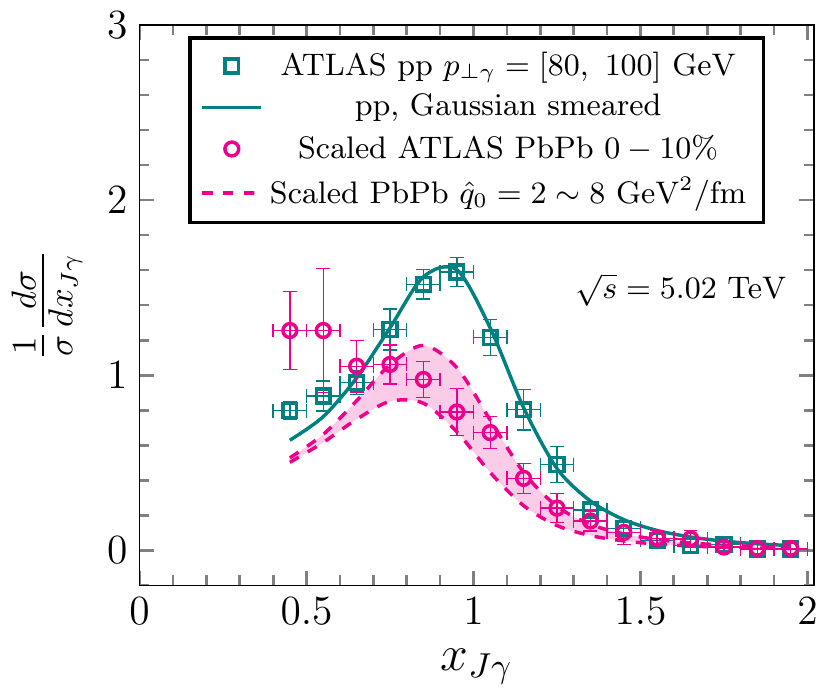}
\includegraphics[width=0.32\linewidth]{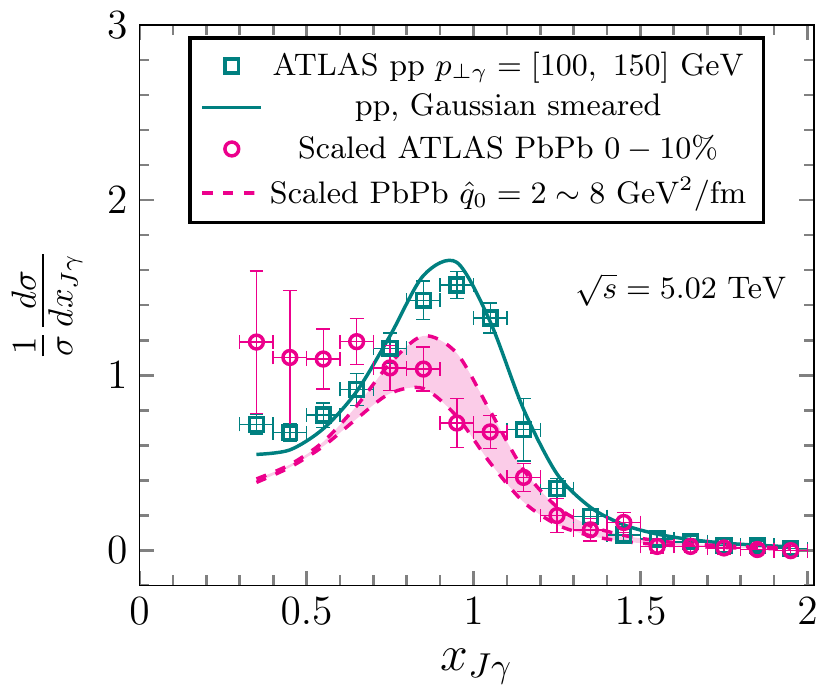}
\end{center}
\caption[*]{Normalized $x_{J\gamma}$ distributions computed for $pp$ collisions and corresponding $x_{J\gamma}$ distributions in $PbPb$ collisions at $5.02$~TeV with $R=0.4$ and jet rapidity cut $|y|<2.1$ compared with the ATLAS data taken from Ref.~\cite{ATLAS:2016tor}. The cuts on the isolated photon is set to be the same as the ATLAS data selection. 
In order to compare with the ATLAS data with the same bin selection, we have removed all events with $x_{J\gamma}$ smaller than the minimal $x_{J\gamma}$ bin chosen in Ref.~\cite{ATLAS:2016tor}. The curves and data for $pp$ collisions are normalized to unity, while the $PbPb$ data are scaled by the same normalization factor as for the $pp$ data to preserve the information of suppression due to energy loss effects. 
}
\label{atlasxj}
\end{figure}

We embed our $pp$ baseline calculation in a realistic modelling of the QGP medium created in heavy ion collisions at the LHC energies. The medium profile includes the collision geometry effect and the space-time evolution of QGP from the OSU (2+1)-dimensional viscous hydrodynamical evolution~\cite{Song:2007ux, Qiu:2011hf}. According to the location of the hard scattering and the momentum orientation of jets, we can compute the energy loss for jets based on the length and temperature of the traversed QGP medium. In addition, we assume the following temperature scaling of $\hat q$ according to dimensional analysis $\hat q = \hat{q}_0 T^3(r, \tau)/T_0^3$, with $T_0=509$ MeV, which corresponds to the temperature in the center of the QGP medium for $0-10\%$ centrality $PbPb$ collisions at $\sqrt{s}=5.02$A TeV. 

Following Ref.~\cite{Baier:2001yt}, the distribution of energy loss in the BDMPS formalism can be approximately written as 
\begin{equation}
\label{Depsilon}
\epsilon D(\epsilon) = \sqrt{ \frac{\alpha^2\omega_c}{2\epsilon}}\exp \left(-\frac{\pi\alpha^2 \omega_c}{2\epsilon}\right).
\end{equation} 
where $\omega_c \equiv  \int \hat{q}_R(\tau) \tau d\tau$ and $\alpha \equiv \frac{2\alpha_s (\mu_r^2) C_R}{\pi}$ with $C_R=C_F (N_C)$ for quark (gluon) jets. $\hat{q}_R$ is defined as the transverse momentum broadening for different species of partons where $\hat{q}_{q}=\hat{q}$ and $\hat{q}_{g}=\frac{C_A}{C_F}\hat{q}$. The strong coupling $\alpha_s$ is set to $0.2$ in Eq.(\ref{Depsilon}) at typical value of $\mu_r^2\sim \hat q L \sim 10~\textrm{GeV}^2$. In reaching this relatively simple formula, the medium induced gluon emission is assumed to be soft. Eq.(\ref{Depsilon}) gives the typical amount of energy loss of $2 \alpha^2\omega_c \sim 4\textrm{GeV}$ if we take medium length $L=2\textrm{fm}$ and $\hat q =6\textrm{GeV}^2/\textrm{fm}$. As long as $2 \alpha^2\omega_c $ is much smaller than the jet energy, the soft gluon approximation should be valid. 

In addition, to more precisely study the jet transport coefficient, we need to distinguish quark jets from gluon jets, since the $\hat{q}$ for gluon jets is $\frac{2N_c^2}{N_c^2-1}$ times of the $\hat{q}$ for quark jet due to larger color factors, which means gluon jets tend to lose more energy than quark jets. In both pQCD and Sudakov resummation calculations, we can separate quark jets from gluon jets and assign different amount of energy loss accordingly. As compared to dijet productions, it is much easier to separate flavour of jets in isolated-photon-jet productions, due to simpler final states.

Based on the smeared $pp$ baseline, we add the energy loss effects in order to describe the data in $PbPb$ collisions, and find that $\hat{q}_0 = 2-8 \textrm{GeV}^2/\textrm{fm}$ at $\sqrt{s}=5.02$A TeV, as shown in Fig.~\ref{atlasxj}. The above extracted range of $\hat{q}_0$ is consistent with $\hat{q}_0 = 2-6 \textrm{GeV}^2/\textrm{fm}$ at $\sqrt{s}=2.76$A TeV from the dijet asymmetries studies\cite{Chen:2016cof}. It is quite reasonable to find that the typical value of $\hat{q}$ becomes slightly larger at higher collisional energy due to roughly $6\%$ increase in the initial temperature from $\sqrt{s}=2.76$A TeV to $\sqrt{s}=5.02$A TeV. We see that the curves and data both show a shift towards smaller value of $x_{J\gamma}$, and an overall suppression of the distribution. This suggests that jets lose a visible amount of energy due to their interaction with the medium which causes the shift. In addition, due to the lower cut on the jet transverse momentum, there are less jets above the cut after passing through the QGP medium, which is the main reason for the suppression. 

\begin{figure}[tbp]
\begin{center}
\includegraphics[width=0.32\linewidth]{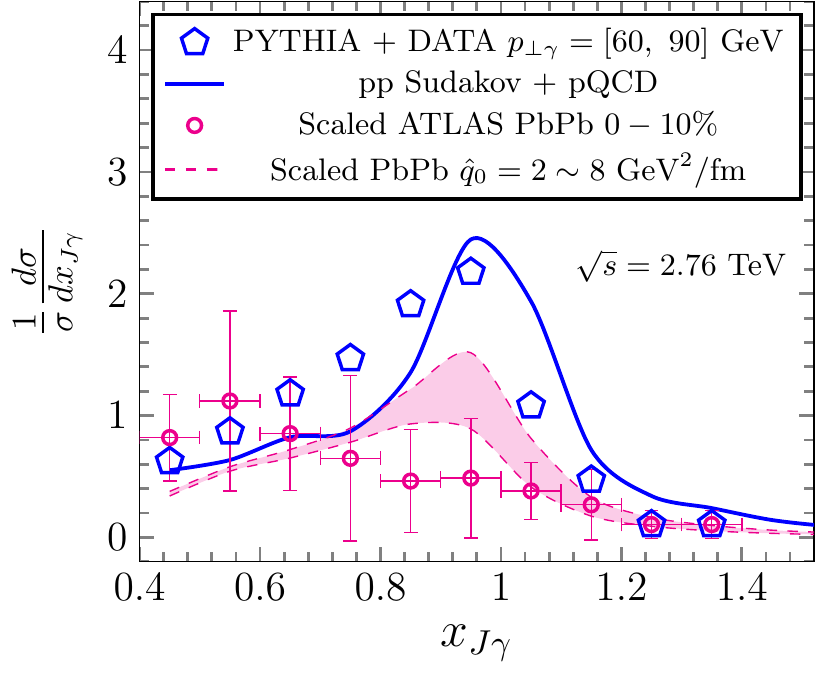}
\end{center}
\caption[*]{Normalized $x_{J\gamma}$ distributions computed for $pp$ collisions and corresponding $x_{J\gamma}$ distributions in $PbPb$ collisions with $0-10\%$ centrality compared with ATLAS data at $\sqrt{s}=2.76$A TeV. The kinematic cuts are $|\eta_\gamma| <1.3$, $|\eta_J| <2.1$, and $|\Delta \phi|> 7\pi/8$, $p_{\perp J} >25$GeV with $R=0.3$. }
\label{atlas-xj-un}
\end{figure}

\begin{figure}[tbp]
\begin{center}
\includegraphics[width=0.32\linewidth]{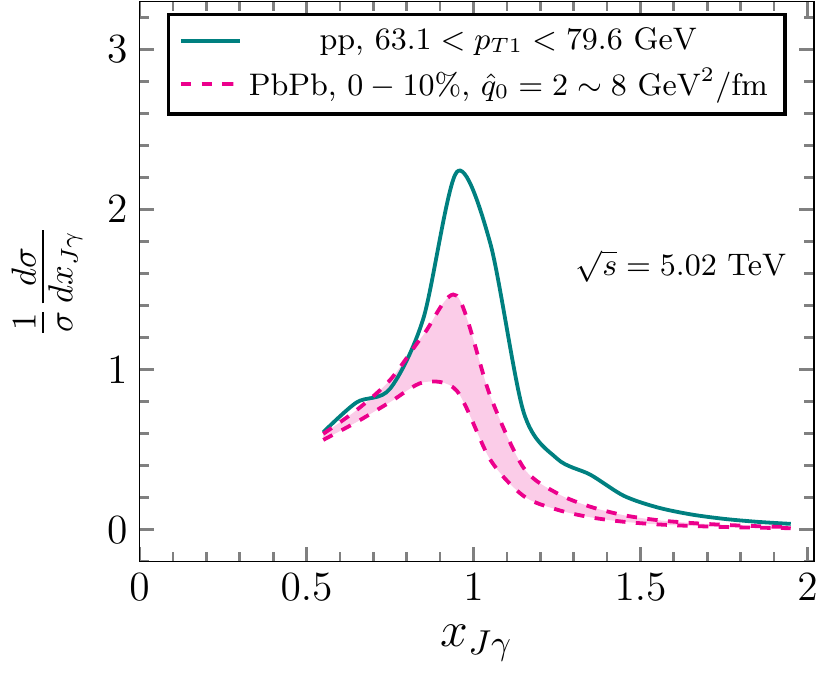}
\includegraphics[width=0.32\linewidth]{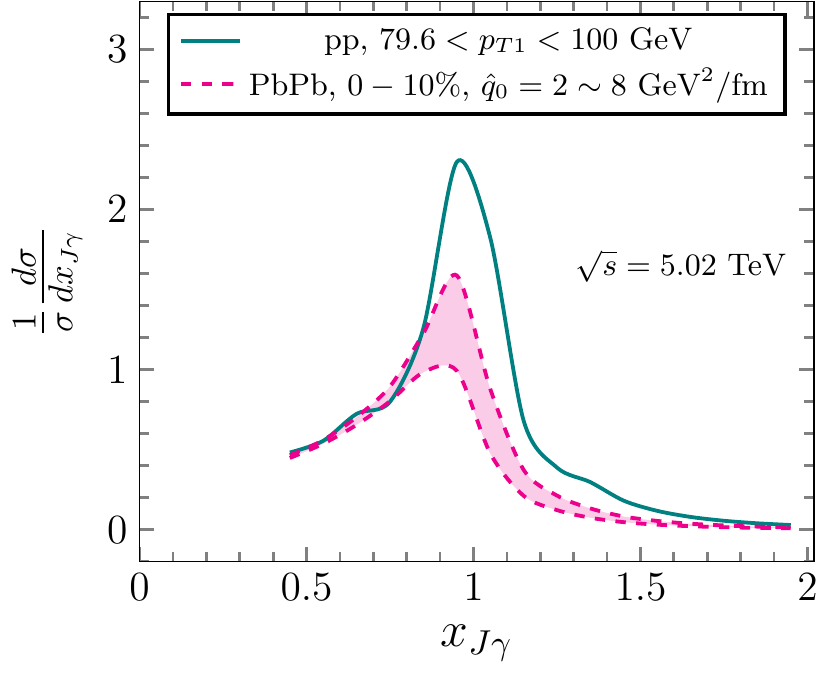} \\
\includegraphics[width=0.32\linewidth]{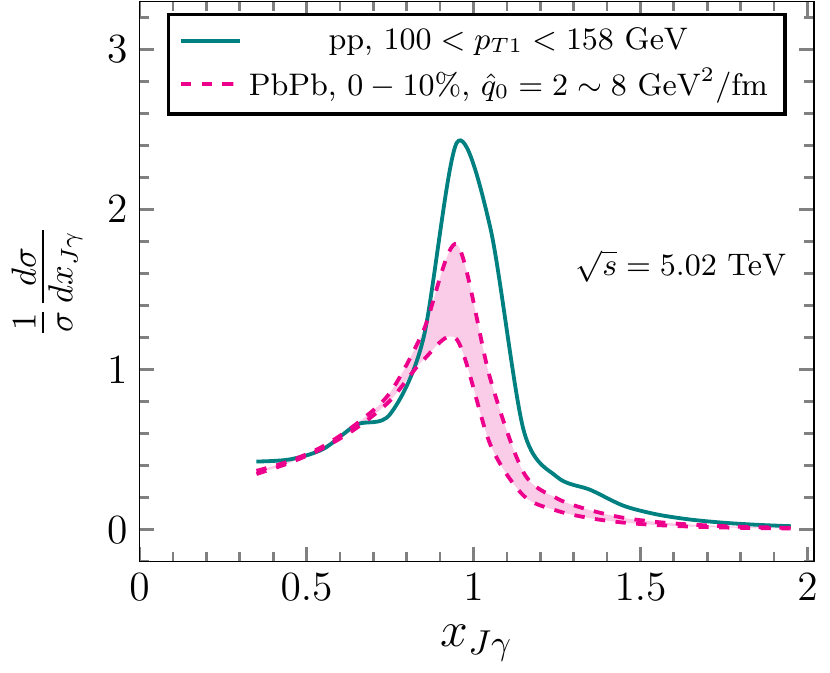}
\includegraphics[width=0.32\linewidth]{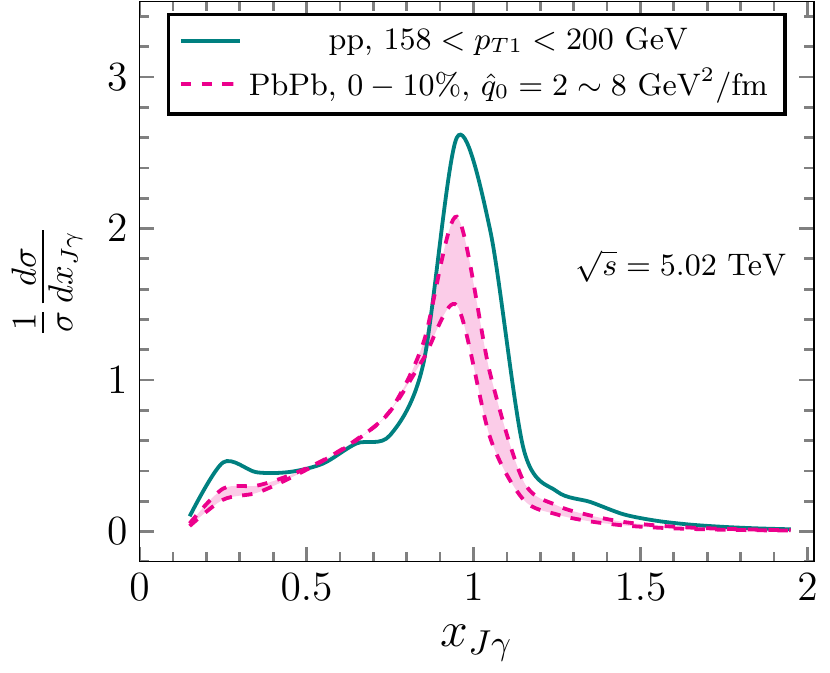}
\end{center}
\caption[*]{Prediction for the normalized $x_{J\gamma}$ distributions computed for $pp$ collisions and corresponding $x_{J\gamma}$ distributions in $PbPb$ collisions at $5.02$ TeV with $R=0.4$ and jet rapidity cut $|y|<2.8$. 
 % computed with the same configurations described in the caption of Fig.~\ref{atlasxj}. 
This prediction should be compared with the fully corrected experimental data.
There seems to be some excess in the small $x_{J\gamma}$ region, which is sensitive to the $p_T$ cut of jets.
}
\label{atlas-xj-prediction}
\end{figure}

In addition, in Fig.~\ref{atlas-xj-un}, we also compare our calculation with the unfolded data from the ATLAS collaboration at  $\sqrt{s}=2.76$A TeV. We find that our results for $pp$ collisions qualitatively agree with the results obtained from PYTHIA plus data overlay. The suppression in the $x_{J\gamma}$ distributions in $PbPb$ collisions also suggests a similar range for the transport coefficient $\hat{q}_0$.

In the end, we would like to make a few comments on the phenomenological study in general. First of all, by comparing the smeared $pp$ and $PbPb$ data, we can see that the $x_{J\gamma}$ distributions suggests that this distribution is shifted towards small $x_{J\gamma}$ and suppressed due to energy loss effects. Second, since the current data still have large error bars, which make it impractical to do $\chi ^2$ fits, future additional and more precise measurement can help us obtain more accurate value of $\hat q$ and understand the energy loss mechanism. Last but not least, we also provide our prediction for the $x_{J\gamma}$ distributions in both $pp$ and $PbPb$ collisions as in Fig. \ref{atlas-xj-prediction}, which can be compared with the unfolded ATLAS data at $\sqrt{s}=5.02$A TeV once it becomes available. The importance of unfolding of data is that it gets rid of a lot of detector effects and make the difference between $pp$ and $PbPb$ collisions more prominent. This can lead us to a more profound and precise understanding of the energy loss in QGP.

%%%%%%%%%%%%%%%%%%%%%%%%%%%%%%%%%%%%%%%%%%%%%%%%%%%%%%%%%%%%%%%%%%%%%%%%%%%%%

\section{Conclusion}

In conclusion, we have calculated the gamma-jet azimuthal angular distribution for both $pp$ and $AA$ collision at LHC kinematics. We find that the pQCD calculation can well describe the experimental data when $\Delta \phi$ is away from $\pi$, while the Sudakov resummation formalism gives a nice agreement with experimental data in the back-to-back region. Furthermore, by extending the so-called Sudakov resummation improved pQCD calculation originally developed for dijet productions, we can also compute the distribution of the isolated-photon and jet momentum imbalance $x_{J\gamma}$. Due to current limitation of data, which have not been fully corrected, we view our results as the predictions based on QCD and energy loss models for future unfolded data. In addition, we employ a Gaussian smearing function to mimic various detector effects, and match to the $pp$ data, which allows us to use it as a baseline to study the energy loss effects in the $PbPb$ collisions. In the end, the $x_{J\gamma}$ distribution allows us to estimate the amount of energy loss due to QGP medium at the LHC energies.

%%%%%%%%%%%%%%%%%%%%%%%%%%%%%%%%%%%%%%%%%%%%%%%%%%%%%%%%%%%%%%%%%%%%%%%%%%%%%

\begin{acknowledgments}
We thank R. Bi, S. Cao, Y.-J. Lee, C. McGinn, D. Perepelitsa, K. Tatar and X. N. Wang for helpful discussions. This material is based on the work supported by the Natural Science Foundation of China (NSFC) under Grant Nos.~11575070, 11775095, 11375072 and 11435004, and the Major State Basic Research Development Program in China under Contract No. 2014CB845400.  This work is also supported by the Agence Nationale de la Recherche under the project ANR-16-CE31-0019.
\end{acknowledgments}


\begin{thebibliography}{99}


\bibitem{Gyulassy:1993hr}
M.~Gyulassy and X.-n. Wang,
\newblock Nucl. Phys. {\bf B420}, 583 (1994), arXiv:nucl-th/9306003.
%%CITATION = NUCL-TH/9306003;%%

\bibitem{Baier:1996kr}
R.~Baier, Y.~L. Dokshitzer, A.~H. Mueller, S.~Peigne, and D.~Schiff,
\newblock Nucl.Phys. {\bf B483}, 291 (1997), arXiv:hep-ph/9607355.
%%CITATION = HEP-PH/9607355;%%

\bibitem{Baier:1996sk}
R.~Baier, Y.~L. Dokshitzer, A.~H. Mueller, S.~Peigne, and D.~Schiff,
\newblock Nucl.Phys. {\bf B484}, 265 (1997), arXiv:hep-ph/9608322.
%%CITATION = HEP-PH/9608322;%%

\bibitem{Baier:1998kq}
R.~Baier, Y.~L. Dokshitzer, A.~H. Mueller, and D.~Schiff,
\newblock Nucl. Phys. {\bf B531}, 403 (1998), arXiv:hep-ph/9804212.
%%CITATION = HEP-PH/9804212;%%

\bibitem{Zakharov:1996fv}
B.~Zakharov,
\newblock JETP Lett. {\bf 63}, 952 (1996), arXiv:hep-ph/9607440.
%%CITATION = HEP-PH/9607440;%%

\bibitem{Gyulassy:1999zd}
M.~Gyulassy, P.~Levai, and I.~Vitev,
\newblock Nucl.Phys. {\bf B571}, 197 (2000), arXiv:hep-ph/9907461.
%%CITATION = HEP-PH/9907461;%%

\bibitem{Wiedemann:2000za}
U.~A. Wiedemann,
\newblock Nucl.Phys. {\bf B588}, 303 (2000), arXiv:hep-ph/0005129.
%%CITATION = HEP-PH/0005129;%%

\bibitem{Arnold:2002ja}
P.~B. Arnold, G.~D. Moore, and L.~G. Yaffe,
\newblock JHEP {\bf 0206}, 030 (2002), arXiv:hep-ph/0204343.
%%CITATION = HEP-PH/0204343;%%

\bibitem{Wang:2001ifa}
X.-N. Wang and X.-f. Guo,
\newblock Nucl. Phys. {\bf A696}, 788 (2001), arXiv:hep-ph/0102230.
%%CITATION = HEP-PH/0102230;%%


\bibitem{Majumder:2010qh}
A.~Majumder and M.~Van~Leeuwen,
\newblock Prog. Part. Nucl. Phys. {\bf 66}, 41 (2011), arXiv:1002.2206.
%%CITATION = ARXIV:1002.2206;%%

\bibitem{Qin:2015srf}
G.-Y. Qin and X.-N. Wang,
\newblock Int. J. Mod. Phys. {\bf E24}, 1530014 (2015), arXiv:1511.00790.
%%CITATION = ARXIV:1511.00790;%%

\bibitem{Blaizot:2015lma}
J.-P. Blaizot and Y.~Mehtar-Tani,
\newblock Int. J. Mod. Phys. {\bf E24}, 1530012 (2015), arXiv:1503.05958.
%%CITATION = ARXIV:1503.05958;%%

\bibitem{Burke:2013yra}
JET, K.~M. Burke {\em et~al.},
\newblock Phys. Rev. {\bf C90}, 014909 (2014), arXiv:1312.5003.
%%CITATION = ARXIV:1312.5003;%%



  

\bibitem{Aad:2010bu}
Atlas Collaboration, G.~Aad {\em et~al.},
\newblock Phys.Rev.Lett. {\bf 105}, 252303 (2010), arXiv:1011.6182.
%%CITATION = ARXIV:1011.6182;%%

\bibitem{Chatrchyan:2011sx}
CMS, S.~Chatrchyan {\em et~al.},
\newblock Phys. Rev. {\bf C84}, 024906 (2011), arXiv:1102.1957.
%%CITATION = ARXIV:1102.1957;%%

\bibitem{Chatrchyan:2012nia}
CMS, S.~Chatrchyan {\em et~al.},
\newblock Phys. Lett. {\bf B712}, 176 (2012), arXiv:1202.5022.
%%CITATION = ARXIV:1202.5022;%%

\bibitem{Qin:2010mn}
G.-Y. Qin and B.~Muller,
\newblock Phys. Rev. Lett. {\bf 106}, 162302 (2011), arXiv:1012.5280,
\newblock [Erratum: Phys. Rev. Lett.108,189904(2012)].
%%CITATION = ARXIV:1012.5280;%%

\bibitem{CasalderreySolana:2010eh}
J.~Casalderrey-Solana, J.~G. Milhano, and U.~A. Wiedemann,
\newblock J.Phys.G {\bf G38}, 035006 (2011), arXiv:1012.0745.
%%CITATION = ARXIV:1012.0745;%%

\bibitem{Young:2011qx}
C.~Young, B.~Schenke, S.~Jeon, and C.~Gale,
\newblock Phys.Rev. {\bf C84}, 024907 (2011), arXiv:1103.5769.
%%CITATION = ARXIV:1103.5769;%%

\bibitem{He:2011pd}
Y.~He, I.~Vitev, and B.-W. Zhang,
\newblock Phys.Lett. {\bf B713}, 224 (2012), arXiv:1105.2566.
%%CITATION = ARXIV:1105.2566;%%

\bibitem{Lokhtin:2011qq}
I.~Lokhtin, A.~Belyaev, and A.~Snigirev,
\newblock Eur.Phys.J. {\bf C71}, 1650 (2011), arXiv:1103.1853.
%%CITATION = ARXIV:1103.1853;%%

\bibitem{ColemanSmith:2012vr}
C.~E. Coleman-Smith and B.~Muller,
\newblock Phys. Rev. {\bf C86}, 054901 (2012), arXiv:1205.6781.
%%CITATION = ARXIV:1205.6781;%%

\bibitem{Renk:2012cb}
T.~Renk,
\newblock Phys. Rev. {\bf C86}, 061901 (2012), arXiv:1204.5572.
%%CITATION = ARXIV:1204.5572;%%

\bibitem{Zapp:2012ak}
K.~C. Zapp, F.~Krauss, and U.~A. Wiedemann,
\newblock JHEP {\bf 03}, 080 (2013), arXiv:1212.1599.
%%CITATION = ARXIV:1212.1599;%%

\bibitem{Ma:2013pha}
G.-L. Ma,
\newblock Phys. Rev. {\bf C87}, 064901 (2013), arXiv:1304.2841.
%%CITATION = ARXIV:1304.2841;%%

\bibitem{Senzel:2013dta}
F.~Senzel, O.~Fochler, J.~Uphoff, Z.~Xu, and C.~Greiner,
\newblock J. Phys. {\bf G42}, 115104 (2015), arXiv:1309.1657.
%%CITATION = ARXIV:1309.1657;%%

\bibitem{Casalderrey-Solana:2014bpa}
J.~Casalderrey-Solana, D.~C. Gulhan, J.~G. Milhano, D.~Pablos, and
  K.~Rajagopal,
\newblock JHEP {\bf 10}, 019 (2014), arXiv:1405.3864,
\newblock [Erratum: JHEP09,175(2015)].
%%CITATION = ARXIV:1405.3864;%%

\bibitem{Ayala:2015jaa}
A.~Ayala, I.~Dominguez, J.~Jalilian-Marian, and M.~E. Tejeda-Yeomans,
\newblock Phys. Rev. {\bf C92}, 044902 (2015), arXiv:1503.06889.
%%CITATION = ARXIV:1503.06889;%%

\bibitem{Milhano:2015mng}
J.~G. Milhano and K.~C. Zapp,
\newblock Eur. Phys. J. {\bf C76}, 288 (2016), arXiv:1512.08107.
%%CITATION = ARXIV:1512.08107;%%

\bibitem{Chang:2016gjp}
N.-B. Chang and G.-Y. Qin,
\newblock Phys. Rev. {\bf C94}, 024902 (2016), arXiv:1603.01920.
%%CITATION = ARXIV:1603.01920;%%

%\cite{Chen:2016cof}
\bibitem{Chen:2016cof} 
  L.~Chen, G.~Y.~Qin, S.~Y.~Wei, B.~W.~Xiao and H.~Z.~Zhang,
  %``Dijet Asymmetry in the Resummation Improved Perturbative QCD Approach,''
  arXiv:1612.04202 [hep-ph].



\bibitem{Wang:1991xy}
X.-N. Wang and M.~Gyulassy,
\newblock Phys.Rev.Lett. {\bf 68}, 1480 (1992).
%%CITATION = PRLTA,68,1480;%%




\bibitem{Banfi:2008qs}
A.~Banfi, M.~Dasgupta, and Y.~Delenda,
\newblock Phys. Lett. {\bf B665}, 86 (2008), arXiv:0804.3786.
%%CITATION = ARXIV:0804.3786;%%


%\cite{Mueller:2012uf}
\bibitem{Mueller:2012uf} 
  A.~H.~Mueller, B.~W.~Xiao and F.~Yuan,
  %``Sudakov Resummation in Small-$x$ Saturation Formalism,''
  Phys.\ Rev.\ Lett.\  {\bf 110}, no. 8, 082301 (2013),  [arXiv:1210.5792 [hep-ph]].
  
\bibitem{Mueller:2013wwa}
A.~H. Mueller, B.-W. Xiao, and F.~Yuan,
\newblock Phys. Rev. {\bf D88}, 114010 (2013), arXiv:1308.2993.
%%CITATION = ARXIV:1308.2993;%%

\bibitem{Sun:2014gfa}
P.~Sun, C.~P. Yuan, and F.~Yuan,
\newblock Phys. Rev. Lett. {\bf 113}, 232001 (2014), arXiv:1405.1105.
%%CITATION = ARXIV:1405.1105;%%

%\cite{Sun:2015doa}
\bibitem{Sun:2015doa} 
  P.~Sun, C.-P.~Yuan and F.~Yuan,
  %``Transverse Momentum Resummation for Dijet Correlation in Hadronic Collisions,''
  Phys.\ Rev.\ D {\bf 92}, no. 9, 094007 (2015)
%  doi:10.1103/PhysRevD.92.094007
  [arXiv:1506.06170 [hep-ph]].
  %%CITATION = doi:10.1103/PhysRevD.92.094007;%%
  
  \bibitem{Mueller:2016gko}
A.~H. Mueller, B.~Wu, B.-W. Xiao, and F.~Yuan,
\newblock Phys. Lett. {\bf B763}, 208 (2016), arXiv:1604.04250.
%%CITATION = ARXIV:1604.04250;%%

\bibitem{Mueller:2016xoc}
A.~H. Mueller, B.~Wu, B.-W. Xiao, and F.~Yuan,
%\newblock (2016), arXiv:1608.07339.
Phys.\ Rev.\ D {\bf 95}, no. 3, 034007 (2017)
 % doi:10.1103/PhysRevD.95.034007
  [arXiv:1608.07339 [hep-ph]].
%%CITATION = ARXIV:1608.07339;%%

\bibitem{Chen:2016vem}
L.~Chen, G.-Y. Qin, S.-Y. Wei, B.-W. Xiao, and H.-Z. Zhang,
%\newblock (2016), arXiv:1607.01932.
Phys.\ Lett.\ B {\bf 773}, 672 (2017)
 % doi:10.1016/j.physletb.2017.09.031
  [arXiv:1607.01932 [hep-ph]].
%%CITATION = ARXIV:1607.01932;%%

 
 \bibitem{Nagy:2001fj}
Z.~Nagy,
\newblock Phys. Rev. Lett. {\bf 88}, 122003 (2002), arXiv:hep-ph/0110315.
%%CITATION = HEP-PH/0110315;%%

\bibitem{Nagy:2003tz}
Z.~Nagy,
\newblock Phys. Rev. {\bf D68}, 094002 (2003), arXiv:hep-ph/0307268.
%%CITATION = HEP-PH/0307268;%%

\bibitem{Abazov:2004hm}
D0, V.~M. Abazov {\em et~al.},
\newblock Phys. Rev. Lett. {\bf 94}, 221801 (2005), arXiv:hep-ex/0409040.
%%CITATION = HEP-EX/0409040;%%

%\cite{Wang:1996yh}
\bibitem{Wang:1996yh} 
  X.~N.~Wang, Z.~Huang and I.~Sarcevic,
  %``Jet quenching in the opposite direction of a tagged photon in high-energy heavy ion collisions,''
  Phys.\ Rev.\ Lett.\  {\bf 77}, 231 (1996)
%  doi:10.1103/PhysRevLett.77.231
  [hep-ph/9605213].
  
  %\cite{Wang:1996pe}
\bibitem{Wang:1996pe} 
  X.~N.~Wang and Z.~Huang,
  %``Study medium induced parton energy loss in gamma + jet events of high-energy heavy ion collisions,''
  Phys.\ Rev.\ C {\bf 55}, 3047 (1997)
  %doi:10.1103/PhysRevC.55.3047
  [hep-ph/9701227].
  
  %\cite{Renk:2006qg}
\bibitem{Renk:2006qg} 
  T.~Renk,
  %``Towards jet tomography: gamma-hadron correlations,''
  Phys.\ Rev.\ C {\bf 74}, 034906 (2006)
 % doi:10.1103/PhysRevC.74.034906
  [hep-ph/0607166].
  
  \bibitem{Zhang:2009rn} 
  H.~Zhang, J.~F.~Owens, E.~Wang and X.~N.~Wang,
  %``Tomography of high-energy nuclear collisions with photon-hadron correlations,''
  Phys.\ Rev.\ Lett.\  {\bf 103}, 032302 (2009)
%  doi:10.1103/PhysRevLett.103.032302
  [arXiv:0902.4000 [nucl-th]].

  
\bibitem{Qin:2009bk} 
  G.~Y.~Qin, J.~Ruppert, C.~Gale, S.~Jeon and G.~D.~Moore,
  %``Jet energy loss, photon production, and photon-hadron correlations at RHIC,''
  Phys.\ Rev.\ C {\bf 80}, 054909 (2009)
%  doi:10.1103/PhysRevC.80.054909
  [arXiv:0906.3280 [hep-ph]].
  
%\cite{Adare:2009vd}
\bibitem{Adare:2009vd} 
  A.~Adare {\it et al.} [PHENIX Collaboration],
  %``Photon-Hadron Jet Correlations in p+p and Au+Au Collisions at s**(1/2) = 200-GeV,''
  Phys.\ Rev.\ C {\bf 80}, 024908 (2009)
  doi:10.1103/PhysRevC.80.024908
  [arXiv:0903.3399 [nucl-ex]].


  \bibitem{Abelev:2009gu} 
  B.~I.~Abelev {\it et al.} [STAR Collaboration],
  %``Studying Parton Energy Loss in Heavy-Ion Collisions via isolated-photon and Charged-Particle Azimuthal Correlations,''
  Phys.\ Rev.\ C {\bf 82}, 034909 (2010)
%  doi:10.1103/PhysRevC.82.034909
  [arXiv:0912.1871 [nucl-ex]].
  
  
  %\cite{Chen:2017zte}
\bibitem{Chen:2017zte} 
  W.~Chen, S.~Cao, T.~Luo, L.~G.~Pang and X.~N.~Wang,
  %``Effects of jet-induced medium excitation in $\gamma$-hadron correlation in A+A collisions,''
  Phys.\ Lett.\ B {\bf 777}, 86 (2018)
 % doi:10.1016/j.physletb.2017.12.015
  [arXiv:1704.03648 [nucl-th]].


  
%
\bibitem{Qin:2012gp} 
  G.~Y.~Qin,
  %``Parton shower evolution in medium and nuclear modification of photon-tagged jets in Pb+Pb collisions at the LHC,''
  Eur.\ Phys.\ J.\ C {\bf 74}, 2959 (2014)
 % doi:10.1140/epjc/s10052-014-2959-3
  [arXiv:1210.6610 [hep-ph]].
  
\bibitem{Wang:2013cia} 
  X.~N.~Wang and Y.~Zhu,
  %``Medium Modification of $\gamma$-jets in High-energy Heavy-ion Collisions,''
  Phys.\ Rev.\ Lett.\  {\bf 111}, no. 6, 062301 (2013)
  %doi:10.1103/PhysRevLett.111.062301
  [arXiv:1302.5874 [hep-ph]].

%\cite{Ma:2013bia}
\bibitem{Ma:2013bia} 
  G.~L.~Ma,
  %``Towards detailed tomography of high energy heavy-ion collisions by $\gamma$-jet,''
  Phys.\ Lett.\ B {\bf 724}, 278 (2013)
%  doi:10.1016/j.physletb.2013.06.029
  [arXiv:1302.5873 [nucl-th]].
  
  %\cite{Luo:2018pto}
\bibitem{Luo:2018pto} 
  T.~Luo, S.~Cao, Y.~He and X.~N.~Wang,
  %``Multiple jets and $\gamma$-jet correlation in high-energy heavy-ion collisions,''
  arXiv:1803.06785 [hep-ph].
  
  %\cite{Kang:2017xnc}
\bibitem{Kang:2017xnc} 
  Z.~B.~Kang, I.~Vitev and H.~Xing,
  %``Vector-boson-tagged jet production in heavy ion collisions at energies available at the CERN Large Hadron Collider,''
  Phys.\ Rev.\ C {\bf 96}, no. 1, 014912 (2017)
  %doi:10.1103/PhysRevC.96.014912
  [arXiv:1702.07276 [hep-ph]].
  
  
\bibitem{Catani:2002ny} 
  S.~Catani, M.~Fontannaz, J.~P.~Guillet and E.~Pilon,
  %``Cross-section of isolated prompt photons in hadron hadron collisions,''
  JHEP {\bf 0205}, 028 (2002)
 % doi:10.1088/1126-6708/2002/05/028
  [hep-ph/0204023].

%\cite{Belghobsi:2009hx}
\bibitem{Belghobsi:2009hx} 
  Z.~Belghobsi, M.~Fontannaz, J.-P.~Guillet, G.~Heinrich, E.~Pilon and M.~Werlen,
  %``Photon - Jet Correlations and Constraints on Fragmentation Functions,''
  Phys.\ Rev.\ D {\bf 79}, 114024 (2009)
%  doi:10.1103/PhysRevD.79.114024
  [arXiv:0903.4834 [hep-ph]].
  
  
  \bibitem{Dai:2012am} 
  W.~Dai, I.~Vitev and B.~W.~Zhang,
  %``Momentum imbalance of isolated photon-tagged jet production at RHIC and LHC,''
  Phys.\ Rev.\ Lett.\  {\bf 110}, no. 14, 142001 (2013)
  %doi:10.1103/PhysRevLett.110.142001
  [arXiv:1207.5177 [hep-ph]].

\bibitem{Alioli:2010xa} 
  S.~Alioli, K.~Hamilton, P.~Nason, C.~Oleari and E.~Re,
  %``Jet pair production in POWHEG,''
  JHEP {\bf 1104}, 081 (2011)
  %doi:10.1007/JHEP04(2011)081
  [arXiv:1012.3380 [hep-ph]].


%
\bibitem{Klasen:2017dsy} 
  M.~Klasen, C.~Klein-Bösing and H.~Poppenborg,
  %``Prompt photon production and photon-jet correlations at the LHC,''
  JHEP {\bf 1803}, 081 (2018)
  %doi:10.1007/JHEP03(2018)081
  [arXiv:1709.04154 [hep-ph]].


\bibitem{Su:2014wpa} 
  P.~Sun, J.~Isaacson, C.-P.~Yuan and F.~Yuan,
  %``Universal Non-perturbative Functions for SIDIS and Drell-Yan Processes,''
  arXiv:1406.3073 [hep-ph].
  
  
%
\bibitem{Prokudin:2015ysa} 
  A.~Prokudin, P.~Sun and F.~Yuan,
  %``Scheme dependence and transverse momentum distribution interpretation of Collins–Soper–Sterman resummation,''
  Phys.\ Lett.\ B {\bf 750}, 533 (2015)
%  doi:10.1016/j.physletb.2015.09.064
  [arXiv:1505.05588 [hep-ph]].

%\cite{Qiu:2000ga}
\bibitem{Qiu:2000ga} 
  J.~w.~Qiu and X.~f.~Zhang,
  %``QCD prediction for heavy boson transverse momentum distributions,''
  Phys.\ Rev.\ Lett.\  {\bf 86}, 2724 (2001)
  %doi:10.1103/PhysRevLett.86.2724
  [hep-ph/0012058].
  
  
  \bibitem{Owens:1986mp} 
  J.~F.~Owens,
  %``Large Momentum Transfer Production of Direct Photons, Jets, and Particles,''
  Rev.\ Mod.\ Phys.\  {\bf 59}, 465 (1987).
%  doi:10.1103/RevModPhys.59.465
  %%CITATION = doi:10.1103/RevModPhys.59.465;%%
  %595 citations counted in INSPIRE as of 02 Feb 2018

  \bibitem{Baer:1989xj} 
  H.~Baer, J.~Ohnemus and J.~F.~Owens,
  %``A Calculation of the Direct Photon Plus Jet Cross-Section in the Next-To-Leading Logarithm Approximation,''
  Phys.\ Lett.\ B {\bf 234}, 127 (1990).
%  doi:10.1016/0370-2693(90)92015-B
  
  \bibitem{Baer:1990ra} 
  H.~Baer, J.~Ohnemus and J.~F.~Owens,
  %``A Next-to-leading Logarithm Calculation of Direct Photon Production,''
  Phys.\ Rev.\ D {\bf 42}, 61 (1990).
%  doi:10.1103/PhysRevD.42.61

  
\bibitem{Chatrchyan:2012gt} 
  S.~Chatrchyan {\it et al.} [CMS Collaboration],
  %``Studies of jet quenching using isolated-photon+jet correlations in PbPb and $pp$ collisions at $\sqrt{s_{NN}}=2.76$ TeV,''
  Phys.\ Lett.\ B {\bf 718}, 773 (2013)
  %doi:10.1016/j.physletb.2012.11.003
  [arXiv:1205.0206 [nucl-ex]].
  
  
  %\cite{CMS:2013oua}
\bibitem{CMS:2013oua} 
  CMS Collaboration [CMS Collaboration],
  %``Study of Isolated photon jet correlation in PbPb and pp collisions at 2.76TeV  and pPb collisions at 5.02TeV,''
  CMS-PAS-HIN-13-006.
  
%\cite{ATLAS:2016tor}
\bibitem{ATLAS:2016tor} 
  The ATLAS collaboration [ATLAS Collaboration],
  %``Study of photon-jet momentum correlations in Pb+Pb and $pp$ collisions at $\sqrt{s_\mathrm{NN}} = 5.02$ TeV with ATLAS,''
  ATLAS-CONF-2016-110.
  
  %\cite{Sirunyan:2017qhf}
\bibitem{Sirunyan:2017qhf} 
  A.~M.~Sirunyan {\it et al.} [CMS Collaboration],
  %``Study of jet quenching with isolated-photon+jet correlations in PbPb and pp collisions at $\sqrt{s_{_{\mathrm{NN}}}} =$ 5.02 TeV,''
  arXiv:1711.09738 [nucl-ex].
  
  

\bibitem{unfold}
ATLAS,
\newblock ATLAS-CONF-2015-052  (2015).
%%CITATION = ATLAS-CONF-2015-052;%%

\bibitem{Perepelitsa:2016zbe}
ATLAS, D.~V. Perepelitsa,
\newblock Nucl. Phys. {\bf A956}, 653 (2016).
%%CITATION = NUPHA,A956,653;%%


\bibitem{ATLAS:2012cna} 
  [ATLAS Collaboration],
  %``Measurement of the correlation of jets with high $p_{T}$ isolated prompt photons in lead-lead collisions at $sqrt{s_{NN}} =2.76}$ TeV with the ATLAS detector at the LHC,''
  ATLAS-CONF-2012-121.
  
  %\cite{Aad:2012ag}
\bibitem{Aad:2012ag} 
  G.~Aad {\it et al.} [ATLAS Collaboration],
  %``Jet energy resolution in proton-proton collisions at $\sqrt{s}=7$ TeV recorded in 2010 with the ATLAS detector,''
  Eur.\ Phys.\ J.\ C {\bf 73}, no. 3, 2306 (2013)
 % doi:10.1140/epjc/s10052-013-2306-0
  [arXiv:1210.6210 [hep-ex]].
  

\bibitem{Song:2007ux}
H.~Song and U.~W. Heinz,
\newblock Phys.Rev. {\bf C77}, 064901 (2008), arXiv:0712.3715.
%%CITATION = ARXIV:0712.3715;%%

\bibitem{Qiu:2011hf}
Z.~Qiu, C.~Shen, and U.~Heinz,
\newblock Phys.Lett. {\bf B707}, 151 (2012), arXiv:1110.3033.
%%CITATION = ARXIV:1110.3033;%%

\bibitem{Baier:2001yt}
R.~Baier, Y.~L. Dokshitzer, A.~H. Mueller, and D.~Schiff,
\newblock JHEP {\bf 09}, 033 (2001), arXiv:hep-ph/0106347.
%%CITATION = HEP-PH/0106347;%%




  \end{thebibliography}
\end{document}